\documentclass[letterpaper,twocolumn,aps,superscriptaddress,floatfix]{revtex4}

\usepackage{graphicx}
\usepackage{dcolumn}
\usepackage{bm}
\usepackage[mathlines]{lineno}
\usepackage[english]{babel}
\usepackage[utf8]{inputenc}

\usepackage{float}
\usepackage{fancyhdr}
\usepackage{algorithm}
\usepackage{algpseudocode}
\usepackage{courier, enumitem, xcolor}         
\usepackage{comment}
\usepackage{lipsum, svg, calc, amsmath, amsfonts, amssymb, amsthm}
\usepackage{ragged2e}
\usepackage[format=plain, singlelinecheck=false, font=footnotesize]{caption}
\usepackage{pdfpages}
\usepackage{bbm} 
\usepackage{adjustbox}
\usepackage{booktabs} 
\usepackage{soul}

\usepackage[
    colorlinks=true,
     citecolor=blue,
    urlcolor=blue
]{hyperref}

\definecolor{colour3}{RGB}{178,55,250} 
\newcounter{noteMCctr} 

\newcounter{noteZLctr} 

\newcounter{noteEMctr} 

\newcounter{noteDRctr} 

\newcounter{noteSYctr} 

\DeclareCaptionJustification{forcedjustified}{\justify}
\begin{document}

\preprint{APS/123-QED}

\title{Gaussian Boson Sampling for Asset Clustering in Statistical Arbitrage Portfolios
}

\author{Dayne Marcus Lopena\footnote{d.lopena24@imperial.ac.uk}}
 \affiliation{Blackett Laboratory, Department of Physics, Imperial College London,
Prince Consort Rd, London, SW7 2AZ, United Kingdom}
 \affiliation{Centre for Quantum Engineering, Science and Technology (QuEST),
Imperial College London, Prince Consort Rd, London, SW7 2AZ, United Kingdom}
 
\author{Daniel Buguks}
\affiliation{HSBC Holdings Plc., 8 Canada Square, London, E14 5HQ, United Kingdom}
\affiliation{Department of Mathematics, University College London, 25 Gordon Street, London, WC1H 0AY, United Kingdom}

\author{Zhenghao Li}
 \affiliation{Blackett Laboratory, Department of Physics, Imperial College London,
Prince Consort Rd, London, SW7 2AZ, United Kingdom}
 \affiliation{Centre for Quantum Engineering, Science and Technology (QuEST),
Imperial College London, Prince Consort Rd, London, SW7 2AZ, United Kingdom}

\author{Ewan Mer}
\affiliation{Blackett Laboratory, Department of Physics, Imperial College London, Prince Consort Rd, London, SW7 2AZ, United Kingdom}
\affiliation{Centre for Quantum Engineering, Science and Technology (QuEST), Imperial College London, Prince Consort Rd, London, SW7 2AZ, United Kingdom}

\author{Shana H. Winston}
 \affiliation{Blackett Laboratory, Department of Physics, Imperial College London,
Prince Consort Rd, London, SW7 2AZ, United Kingdom}
 \affiliation{Centre for Quantum Engineering, Science and Technology (QuEST),
Imperial College London, Prince Consort Rd, London, SW7 2AZ, United Kingdom}

\author{Shang Yu}
\affiliation{Blackett Laboratory, Department of Physics, Imperial College London, Prince Consort Rd, London, SW7 2AZ, United Kingdom}
\affiliation{Centre for Quantum Engineering, Science and Technology (QuEST), Imperial College London, Prince Consort Rd, London, SW7 2AZ, United Kingdom}

\author{Mihai Cucuringu}
 \affiliation{Department of Mathematics, University of California, Los Angeles, California, USA}
 \affiliation{Department of Statistics, University of Oxford, Oxford, Oxfordshire, United Kingdom}
\affiliation{Oxford-Man Institute of Quantitative
Finance, University of Oxford, Oxford, Oxfordshire, United Kingdom}

\author{Del Rajan\footnote{del.rajan@hsbc.com}}
\affiliation{HSBC Holdings Plc., 8 Canada Square, London, E14 5HQ, United Kingdom}

\author{Philip Intallura}
\affiliation{HSBC Holdings Plc., 8 Canada Square, London, E14 5HQ, United Kingdom}

\author{Raj B. Patel\footnote{raj.patel1@imperial.ac.uk}}
 \affiliation{Blackett Laboratory, Department of Physics, Imperial College London,
Prince Consort Rd, London, SW7 2AZ, United Kingdom}
 \affiliation{Centre for Quantum Engineering, Science and Technology (QuEST),
Imperial College London, Prince Consort Rd, London, SW7 2AZ, United Kingdom}

\begin{abstract}

Gaussian Boson Sampling (GBS) provides a native photonic quantum heuristic for sampling dense subgraphs from adjacency matrices, offering a scalable physical approach to combinatorial graph search problems. Simultaneously, correlation matrix clustering algorithms, such as Spectral and SPONGE, have established robust benchmarks for identifying co-moving assets from correlation matrices in statistical arbitrage (StatArb) strategies. In this work, we map S\&P 500 residual correlation data into GBS-compatible adjacency matrices. We benchmark those classical clustering algorithms against two quantum clustering algorithms, GBS Boost and our novel GBS Roots, to construct dynamic, market-neutral portfolios over a rolling one-year window. Simulations across distinct macroeconomic regimes reveal that quantum clustering generates superior alpha within large stock universes during periods of high volatility, effectively isolating structural market idiosyncrasies. Crucially, this economic advantage persists under simulated low-loss conditions and extends into high-loss regimes via the application of coherent displacement to compensate for photon loss. Our findings underscore the efficacy of GBS-derived graph clustering in constructing robust StatArb portfolios, establishing a quantum foundation for broader quantitative finance applications.

\end{abstract}
\maketitle

\section{Introduction}


The intersection of quantum computing and quantitative finance has primarily focused on gate-based architectures~\cite{stamatopoulos2020option}, quantum walks~\cite{slate2021quantum, chang2025quantum} or quantum annealers to solve portfolio optimisation problems~\cite{orus2019quantum, rosenberg2015solving}. However, a distinct paradigm of photonic quantum computing, Gaussian Boson Sampling (GBS), provides a complementary approach to quantum computation. In GBS~\cite{aaronson2011computational, hamilton2017gaussian}, single-mode squeezed vacuum states interfere in a linear interferometer composed of beamsplitters and phase shifters to generate output probability distributions that are believed to be classically intractable to simulate. 

Recent experiments ~\cite{Jiuzhang1.0,Jiuzhang2.0,Borealis,Jiuzhang3.0,liu2026gaussian} have 
scaled GBS to regimes that challenge, and in some settings surpass, the best known classical algorithms~\cite{oh2022classical,oh2024classical,liu2023simulating,zhang2025efficient} for this sampling task. These quantum advantage demonstrations rely heavily on reducing the circuit connectivity or programmability to mitigate photon loss, which would otherwise render these experiments efficiently classically simulable. While suitable for demonstrating quantum advantage, these experiments offer 
limited practical applicability in real-world applications.


Nevertheless, GBS has been explored for a variety of applications. These include enhanced image recognition~\cite{gong2025enhanced}, quantum machine learning~\cite{zhugbs, montesinos2026benchmarking}, vibronic spectra simulation~\cite{huh2015boson}, molecular docking~\cite{banchi2020molecular} and fraud detection~\cite{he2025time}. Additionally, experiments have been performed with fully programmable universal GBS devices to address max-clique search problems arising in drug discovery~\cite{yu2023universal}, and topological network analysis~\cite{yu2025babbage}. All of these applications seek to demonstrate empirical quantum~\cite{tn89-g1xz} advantage through GBS, while simultaneously targeting practically relevant computational problems.


The applications of GBS in Ref.~\cite{bradler2018gaussian, arrazola2018using, schuld2020measuring, banchi2020molecular, yu2023universal,yu2025babbage,uhlenbeck1930theory} are particularly interesting as they highlight  an intrinsic connection between GBS and graphs. GBS is able to preferentially sample dense subgraphs, and these samples can be leveraged to solve graph problems such as the maximum-weighted clique search~\cite{banchi2020molecular}, maximum hafnian problem~\cite{deng2023solving}, graph isomorphism problem~\cite{bradler2018graph}, densest $k$-subgraph problem~\cite{deng2023solving, arrazola2018using}, and graph colouring~\cite{epequin2026quantumphotonicapproachgraph}. 

In this work, we empirically demonstrate the near-term potential of photonic quantum computing, via GBS, to construct 
graph-based statistical arbitrage (StatArb) trading portfolios. 
We adapt GBS Boost~\cite{bonaldi2024boost} and propose the novel GBS Roots graph clustering algorithm, that aim to maximise the sum of intra-cluster edge weights, producing stronger statistical arbitrage opportunities in our portfolio. We benchmark these quantum algorithms against classical clustering algorithms, Spectral and signed positive over negative generalised eigenproblem (SPONGE), and highlight their economic advantage under low photon loss, and for larger asset universes and during periods of elevated market volatility.

Demonstrating quantum advantage with GBS faces two primary challenges: i) experimental limitations such as photon loss and distinguishability, and ii) spoofing, for instance, when input matrices possess classically exploitable structures like sparsity or non-negativity. Our demonstration of empirical quantum advantage relies on the efficacy of GBS as a heuristic solver rather than formal complexity-theoretic proofs. Through simulations on large correlation graphs under the assumption of zero photon loss, we demonstrate that quantum algorithms achieve superior total returns relative to both classical Spectral and SPONGE algorithms. However, performance degrades and becomes increasingly variable under photon loss, and our restriction to a 10--20\% subset of the broader stock universe inherently misses some market correlations.

However, when coherent displacement is introduced in the presence of loss to stabilise mean photon numbers, we find that performance recovers. Moreover, this economic advantage is prevalent when the sample period contains a volatile market. In the presence of experimental imperfections, an empirical exponential quantum advantage is not expected. Nevertheless, industries like finance still find empirical polynomial quantum advantages valuable, even in the presence of such imperfections.

\begin{figure*}[t!] 
\centering
{\includegraphics{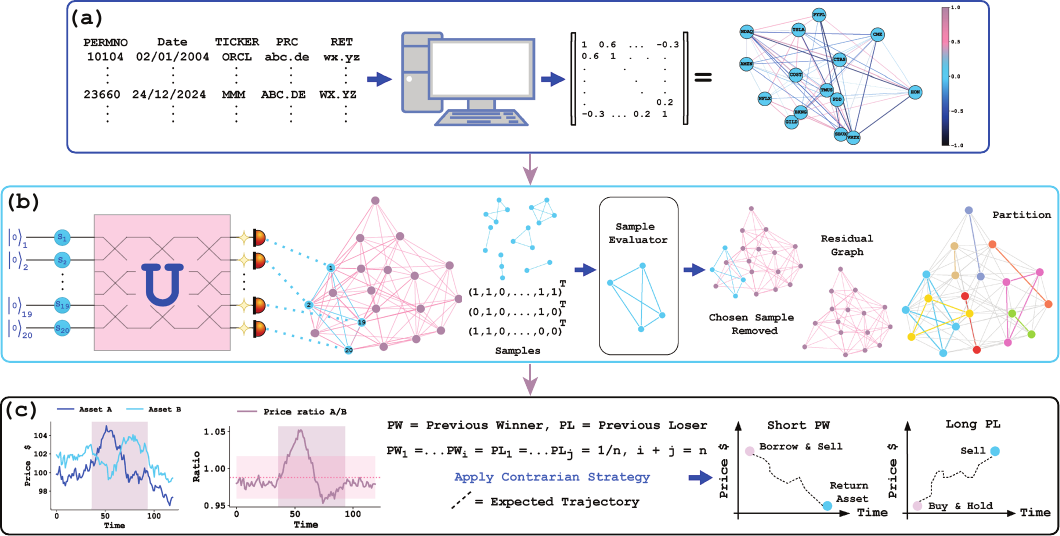}}
\caption{Overview of the GBS-driven StatArb framework. (a) Stock data is processed to construct a residual-return correlation matrix. Zeroing the main diagonal yields the adjacency matrix of the correlation graph. (b) The GBS clustering procedure. The correlation graph is encoded into a GBS device through its adjacency matrix. The device generates subgraph samples, which are evaluated against an edge-weight criterion to select the highest-scoring cluster. This selected subgraph is then removed from the graph to produce a residual graph, and the process is iteratively applied to generate a complete partition of the correlation network. (c) Application to StatArb. The initial graphs illustrate how temporary price anomalies arise among co-moving stocks and the threshold at which these deviations become significant. By evaluating arbitrage signals within the partitioned clusters, constituent stocks are classified as previous winners or losers and assigned uniform portfolio weights. A contrarian trading strategy is subsequently executed to capture returns.}
\label{fig:main}
\end{figure*}

\section{Strategy Overview} \label{sec:overview}

We provide an overview of the main 
components of our
GBS application into statistical arbitrage (StatArb). A workflow diagram can be seen in Figure \ref{fig:main}. The data processing step to construct the correlation matrix needed for the following GBS clustering and StatArb steps is shown in Figure \ref{fig:main}(a). More details on the StatArb part can be found in Appendices \ref{app:stat_arb_strat}, \ref{app:eval_spacings} and for the GBS part see Section \ref{sec:gbs_clus} and Appendices \ref{app:gbs_theory}, \ref{app:pre-pro}, and \ref{app:dss_with_gbs}.

\subsection{Subgraphs from GBS for Clustering}
The general GBS clustering procedure is visualised in Figure \ref{fig:main}(b); it involves mapping a graph (with a corresponding adjacency matrix) onto a GBS device, which performs a matrix decomposition that recovers the squeezing values $s_k$ and unitary $\mathbf{U}$, unique to the input adjacency matrix. Then, the GBS device will interfere these squeezed photons and detect them to produce a photon output statistic, based on which modes the photons were detected. These detected modes correspond to a subset of nodes of the mapped graph, thus producing a subgraph sample. The probability of measuring a specific photon statistic is proportional to the hafnian, a computationally hard matrix function, of the corresponding submatrix. Crucially, this probability is biased towards dense subgraphs, due to the connection between graph connectivity and perfect matchings, which the hafnian counts when the input is an adjacency matrix.

Running this GBS device many times produces many dense subgraph samples that we feed into a sample evaluator that selects the sample with the best property, e.g., graph density. This sample is chosen as the cluster that is removed from the input graph, leaving us with a residual graph. We iterate this process by inputting the adjacency matrix of the residual graph into the GBS device, until we get a partition of the original correlation graph. 

This framework for clustering StatArb portfolios using GBS, offers an alternative to   classical graph clustering methods~\cite{jin2023correlation}. We adapt an existing algorithm into GBS Boost~\cite{bonaldi2024boost} and propose a novel agglomerative-style method, GBS Roots. Using S\&P 500 stock price data~\cite{wrds}, we benchmark these hybrid algorithms against classical Spectral~\cite{ng2001spectral} and SPONGE~\cite{cucuringu2019sponge} methods. We also compare our approach to a random clustering baseline (under photon loss) and a classical GBS method, QIC-GBS~\cite{Oh2024quantum-inspired}. Throughout this work, ``quantum methods'' refer specifically to GBS Boost and GBS Roots, while ``GBS methods'' additionally includes QIC-GBS.

The ability to sample interconnected communities suggests that GBS can serve as a heuristic for clustering problems. Using edge weights, we frame the task as a correlation clustering problem~\cite{bansal2004correlation} applied to financial correlation graphs (FCGs). In FCGs, nodes represent stocks, and edges denote their pairwise correlations (relative price movements), characterised by a correlation matrix. To obtain a (weighted) adjacency matrix structure, we zero the main diagonal of a correlation matrix. 


\subsection{Statistical Arbitrage \& Trading Portfolio Construction}

Quantitative trading strategies utilise mathematical models to identify and exploit predictable price movements in financial assets. A prominent example is Statistical Arbitrage (StatArb), which relies fundamentally on mean reversion. This strategy assumes that price deviations, whether from an asset's historical average or relative to highly correlated assets, are temporary anomalies that will tend to revert to their mean. These deviations can persist for days or resolve in sub-second intervals, bordering on high-frequency trading~\cite{treleaven2013algorithmic}. Profit is generated by exploiting these transient inefficiencies; trading systems simultaneously take offsetting positions, buying the underpriced asset and short-selling the overpriced one, to capture returns upon convergence. The idea of our StatArb strategy can be seen in Figure \ref{fig:main}(c).

Accurately clustering correlated assets facilitates StatArb. StatArb strategies range from pairs trading~\cite{elliott2005pairs} to multi-asset group strategies using principal component analysis (PCA)~\cite{avellaneda2010statistical}.  While this core tenet remains consistent, StatArb implementations vary in their asset similarity metrics, including distance, co-integration, and time series analysis~\cite{krauss2017statistical}. Motivated by Ref.~\cite{jin2023correlation}, we apply our GBS clustering framework to construct StatArb portfolios. Their work pioneered grouping co-moving assets based on correlations to assess the resulting economic value of the clusters. 

Our methodology diverges by substituting classical clustering algorithms with quantum alternatives. We evaluate the performance of these quantum methods against two of the highest-performing classical algorithms from~\cite{jin2023correlation} (SPONGE and Spectral), alongside a classical GBS method adapted from~\cite{Oh2024quantum-inspired}. This comparative analysis is evaluated using financial indicators, such as total return and the Sharpe ratio, as well as graph-theoretic metrics such as weighted density (WD).

We construct a zero-cost, mean-reverting trading portfolio across rolling windows (detailed in Appendix \ref{app:stat_arb_strat}). In each window, an $N \times N$ correlation matrix derived from the preceding $p$ days is clustered to isolate co-moving assets by maximising intra-cluster correlations. Within these partitions, stocks are classified as previous winners or losers relative to the cluster's mean return. We execute an equally weighted contrarian strategy by shorting winners and taking long positions in losers, held for $l$ days before recalculating the returns, rebalancing and rolling the window forward.

For classical algorithms requiring a predefined target cluster count $K$, we introduce a data-driven selection criterion. For a correlation matrix $\mathbf{\mathcal{C}}$ with ordered eigenvalues $\lambda_i$, $K$ is defined as the number of eigenvalue spacings ($s_i := \lambda_i - \lambda_{i+1}$) that exceed a predefined threshold (Appendix \ref{app:eval_spacings}). In contrast, the GBS-based algorithms bypass this prerequisite entirely, allowing the quantum heuristic to partition the graph without requiring a predefined number of clusters.

\section{gbs clustering} \label{sec:gbs_clus}

GBS is frequently deployed as an advanced heuristic to initialise graph-search algorithms for graph problems, such as densest-$k$ subgraph searches~\cite{deng2023solving, bradler2018gaussian}, max-clique identification~\cite{banchi2020molecular,yu2023universal,yu2025babbage,yu2026clavina}, and network clustering~\cite{bonaldi2024boost}. Abstractly, GBS biases a graph search toward optimal regions of a solution space. This process functions analogously to a quasi-warm-start optimisation, a technique already introduced in quantum computing~\cite{Egger2021WarmStart} and further explored via adaptive bias in a Quantum Approximate Optimisation Algorithm (QAOA) for maximum cut problems~\cite{Yu2025WSabQAOA}.

To contextualise our financial application, it is instructive to examine how the objective shifts across domains. For instance, addressing the densest $k$-subgraph problem~\cite{arrazola2018using} represents a direct application of GBS, leveraging its inherent tendency to sample dense subgraphs. Conversely, applying GBS to molecular docking~\cite{banchi2020molecular, yu2023universal} or topological network analysis~\cite{yu2025babbage} shifts the objective to solving a maximum weighted clique problem. 


In our financial setting, however, simply using GBS to identify dense subgraphs is insufficient. Applying it purely as a structural heuristic is insufficient, as there is no intrinsic link between the output of the GBS device and the ultimate profit and loss (PnL) generated by the StatArb portfolio. Maximising traditional graph-theoretic properties, such as connectivity, does not necessarily translate into improved financial performance. Therefore, we define our specific graph problem as finding a partition of the correlation graph, $\mathbf{\mathcal{C}}$, that maximises intra-cluster correlations. We formalise this objective with a value function that sums the correlations across all clusters in a given partition:

\begin{align} \label{value_func}
    \text{Value}(C) := \sum_{\alpha} \sum_{u,v \in C_{\alpha}} w_{uv},
\end{align}

\noindent where $C = \{C_{\alpha}\}_{\alpha=1}^N$ is a clustering (a partition of our correlation graph into clusters $C_{\alpha}$), with $u$ and $v$ denoting stocks, and $w_{uv}$ representing the pairwise correlation between those stocks.

\subsection{Correlation Clustering}

The correlation clustering framework introduced in Ref.~\cite{bansal2004correlation} is known to be NP-hard. While quantum computing is not expected to solve this problem class efficiently in deterministic polynomial time, we hypothesise that specialised quantum hardware can generate high-quality approximate solutions more effectively than classical alternatives.

Standard variations of correlation clustering focus on minimising a cost function associated with edge disagreements~\cite{shakiba2025correlation, davies2023fast}, whereas our framework tackles a max-agree variant aimed at maximising intra-cluster correlations. A relevant precedent is the GCS-Q algorithm~\cite{macaluso2025quantum}, which utilises quantum annealing to address a similar max-agree problem. By formulating a quadratic unconstrained binary optimisation (QUBO) objective and building upon prior work in quantum graph coalition structure generation~\cite{venkatesh2023gcs}, they evaluate candidate subgraphs using a cut condition based on the sum of subgraph edge weights. This approach directly motivates our cluster evaluation metric defined in Equation~(\ref{value_func}).

Typical GBS graph search algorithms are benchmarked by success probabilities or the number of samples required to identify a dense subgraph. However, because financial correlation graphs are inherently fully connected, we bypass these conventional metrics. Instead, our primary focus is the downstream utility of these dense subgraphs in constructing high-quality partitions for trading strategies. Consequently, our Max-Agree Correlation Clustering Problem (MACCP) explicitly extends beyond the standard graph search tasks for which GBS is typically employed.



Unlike quantum algorithms that minimise global cost functions~\cite{willsch2020benchmarking, bapst2013quantum}, our value function provides a global quality metric for the partition, establishing a direct bridge between GBS and StatArb. Since StatArb exploits temporal price deviations among co-moving assets, dominant trading signals require partitions with strong positive intra-cluster correlations, so there is confidence that the arbitrage signals are genuine and not stochastic noise. Increasing this partition value is expected to increase the likelihood for clustering algorithms to produce economic returns. 

Although highly optimised classical subroutines could rigorously maximise the MACCP objective using GBS samples, this creates an attribution problem by obscuring the empirical quantum advantage. To address this, GBS Roots (and adapted GBS Boost) act as lightweight, greedy heuristics. By restricting classical post-processing to simple density evaluation and sequential union-building, we aim to ensure that any downstream economic outperformance can be more directly attributed to the underlying quantum sampling.

\begin{figure*}[t!] 
\includegraphics[width=1\linewidth,height=1\textheight,keepaspectratio]{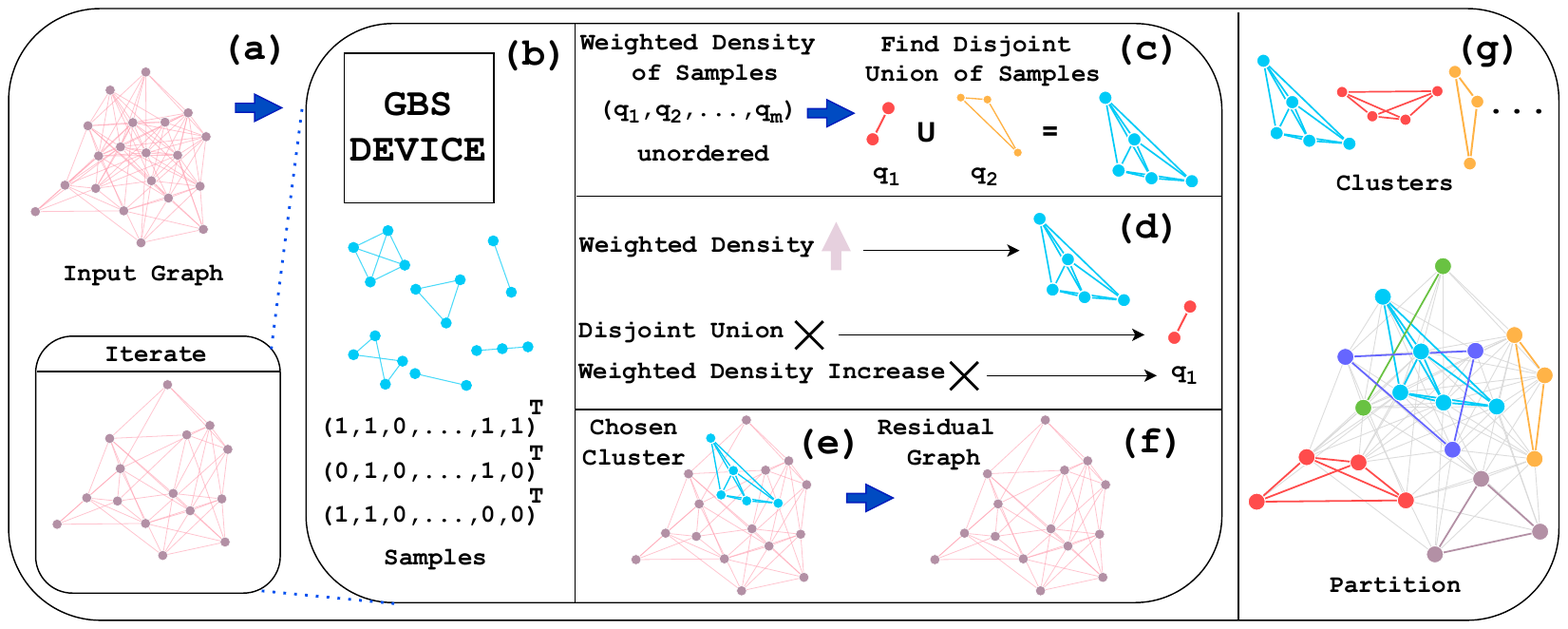}
\vspace*{-6.2cm}
\caption{Overview of the GBS Roots process. (\textbf{a}) An initial correlation graph $\mathcal{G}$, with specific correlations and stock labels omitted for clarity. (\textbf{b}) The correlation graph is sampled using a GBS device to generate candidate subgraphs. (\textbf{c}) The weighted density of the samples is calculated. A cluster $\mathcal{U} = \bigcup_\alpha \tilde{q}_{\alpha}$ is formed by iteratively expanding a disjoint union of these samples, ensuring all $\tilde{q}_{\alpha}$ remain mutually disjoint while recording the utilised nodes. (\textbf{d}) This union expansion terminates if for the next sample in the queue $q_{i+1}$, the union would no longer be disjoint, and the total WD does not increase. (\textbf{e}) Removing the finalised cluster yields the residual graph $\mathcal{G} \setminus \mathcal{U}$. (\textbf{f}) This residual graph is fed back into the GBS device to generate the next cluster. (\textbf{g}) The cycle ((\textbf{b})--(\textbf{f})) repeats until the residual graph contains only two nodes, which are then assigned as the final cluster. If a single node remains at iteration $f+1$, the residual graph from iteration $f$ is retroactively designated as the final cluster. The final output is a complete partition of the input graph.}
\label{fig:roots_process}
\end{figure*}

\subsection{GBS Roots}

We employ clustering methods to group stocks with similar features and correlation structures so that the intra-cluster correlation is maximised. Common classical clustering methods used in this space are referred to as spectral methods, as these clustering algorithms operate on the eigen-spectrum of matrix operators constructed from the data. Two such spectral methods we will use in our benchmarking are Spectral clustering \cite{ng2001spectral} and SPONGE (the symmetric variant in particular) clustering \cite{cucuringu2019sponge, cucuringu2021regularized}. Both methods generally determine some number of the smallest eigenvectors of a variant of a Laplacian matrix and then perform $K$-means$++$ clustering to get the clusters.

Although our value metric is biased toward larger clusters, we can normalise it by the maximum potential value to compute the WD, which we use to guide cluster selection at each iteration of clustering. We define the WD of a cluster $\mathcal{C}$ as:


\begin{align} \label{eqn:graph_weighted_density}
    \text{WD}(\mathcal{C}) := \frac{2\sum_{ij}\mathcal{C}_{ij}}{|V|(|V| - 1)}.
\end{align}

GBS Roots is an agglomerative clustering algorithm that constructs a disjoint union of GBS samples at each iteration for its cluster. It uniquely employs WD for cluster selection, and builds a partition by systematically aggregating clusters.

\begin{algorithm}[t!]
\caption{GBS Roots} \label{alg:roots}
\begin{algorithmic}[1]
\Require Correlation matrix $\mathbf{\mathcal{C}}$ with graph $\mathcal{G}$, mean photon number $\bar{n}$, number of stocks $N$
\Ensure $C = \{C_1, C_2, \hdots\}$, a partition of $\mathcal{G}$
\State $C \gets \emptyset$
\State $\mathbf{\mathcal{A}} \gets \text{max}(0,\mathbf{\mathcal{C}} - \mathbf{I})$ 
\While{$\mathbf{\mathcal{A}} \in \mathbb{R}_{[0,1]}^{N \times N}, \; N > 2$} 

    \State $P \gets \texttt{GBS}(\mathcal{A}, \bar{n}, N)$ \Comment{Generate GBS samples}
    \State $q_1 \gets \arg\max_{p \in P} \text{WD}(p)$ \Comment{Highest WD sample}
    \State $Q \gets (q_1, p\in P\setminus q_1)$ \Comment{Unordered sequence}
    \State $\mathcal{U}\gets \emptyset, \text{used} \gets \emptyset$
    \For{$q \in Q$} \Comment{Greedy multi-packing}

        \If{$q \cap \text{used} \neq \emptyset$ and $(\text{WD}(\mathcal{U} \cup q) \leq \text{WD}(\mathcal{U}))$}
            \State \textbf{break} \Comment{Stop if no overlap or no WD gain}
        \EndIf
            \State $\mathcal{U} \gets \mathcal{U} \cup \{q\}$ \Comment{Add $q$ if no overlap}
            \State $\text{used} \gets \text{used} \cup q$ \Comment{Mark stocks in $q$ as taken}
    \EndFor
    \If{$N - |\text{used}| = 1$} \Comment{Prevent singletons}
        \State $C \gets C \cup \mathcal{G}$
        \State \textbf{break}
    \Else
        \State $C \gets C \cup \mathcal{U}$
        \State $\mathbf{\mathcal{A}} \gets \text{reduce}(\mathbf{\mathcal{A}}, \mathcal{U})$
        \State $\bar{n}, N \gets \text{update}(\mathcal{A})$
    \EndIf
\EndWhile
\If{$N = 2$} \State $C \gets C \cup \mathcal{G}$ \Comment{Take residual graph as a cluster} 
\EndIf
\State \textbf{return} $C$
\end{algorithmic}
\end{algorithm}

The GBS Roots algorithm (Algorithm \ref{alg:roots}) initialises an empty set to store the chosen clusters at each iteration. It prepares the input adjacency matrix $\mathbf{\mathcal{A}}$ of its corresponding graph $\mathcal{G}$ for the GBS process (detailed in Appendix \ref{app:dss_with_gbs}). Unlike classical methods, GBS Roots does not require a predefined cluster count $K$; the iterative process continues until the graph cannot be further reduced.

Each iteration begins by drawing a batch of $H$ GBS samples, calculating their weighted densities, and organising them into an unordered set, with the highest-WD sample is placed first. We then initiate a greedy multi-packing routine. Starting with $q_1$, we check the intersection with the next sample, $q_2$. If they are disjoint and their union increases the total WD, we merge them and mark their nodes as used. This evaluation iterates through the remaining unordered samples $q_j$, continuously updating the union and used sets, until the union is no longer disjoint or the WD no longer increases. The resulting disjoint union $\mathcal{U}$ forms a cluster, and its constituent nodes are removed from $\mathcal{G}$ to create the residual graph $\mathcal{G}\setminus\mathcal{U}$. The adjacency matrix $\mathbf{\mathcal{A}}$ is updated to reflect the reduction in the number of nodes $N$ and the mean photon number $\bar{n}$. Feeding this residual graph back into the GBS device triggers the next iteration. This Roots process, illustrated in Figure \ref{fig:roots_process}, outputs a set of clusters $C$ that partitions the original graph $\mathcal{G}$.

Although GBS Roots does not explicitly maximise the absolute value metric per cluster, the final global clustering is still guided by this objective. Furthermore, GBS Roots tends to produce larger clusters, which inherently increases intra-cluster diversification and mitigates idiosyncratic risk. 

Assuming equal weighting, the variance of an $n$-stock cluster is $\sigma^2 = \frac{1}{n^2} \sum_i^n \sigma_i^2 + \frac{1}{n^2} \sum_{i \neq j} \sigma_{ij}^2$. 
As cluster size $n$ grows, 
the contribution of idiosyncratic volatility is expected to diminish relative to the common co-movement component. This stability is expected to yield steadier trading signals, reduced turnover between previous winners and losers, and ultimately, smoother PnL trajectories. Consequently, simulations involving smaller stock universes (as seen in Appendix \ref{app:mos}) are expected to exhibit higher variance in returns. See Appendix \ref{app:gbs_boost} for details on the GBS Boost algorithm. 
\section{Benchmarking} \label{sec:bench}




We evaluate the economic and clustering performance of StatArb portfolios using GBS Boost and GBS Roots, benchmarked against two classical algorithms (Spectral and SPONGE) and a quantum-inspired classical GBS (QIC-GBS) algorithm \cite{Oh2024quantum-inspired}. QIC-GBS efficiently approximates the GBS distribution for non-negative input matrices using an independent pairs and singles distribution. By sampling edges via a Poisson distribution defined by the adjacency matrix weights, QIC-GBS effectively implements the GBS Boost clustering logic (Algorithm \ref{alg:boost}) using a classical engine.

Appendices \ref{app:stat_arb_strat}, \ref{app:data_api}, and \ref{app:metrics} detail the StatArb strategy parameters (including threshold values), data sources, and evaluation metrics, respectively. For details on the simulations, see Appendix \ref{app:params}, \ref{app:data_api}. Across all benchmarks, the stock universe size is adjusted solely to accommodate computational constraints.

\subsection{Large Stock Universe without Loss}

\begin{figure*}[t!]
    \centering
    \adjustbox{trim=0 0 0 1.6cm, clip}{\includesvg[width=\linewidth,height=\textheight,keepaspectratio]{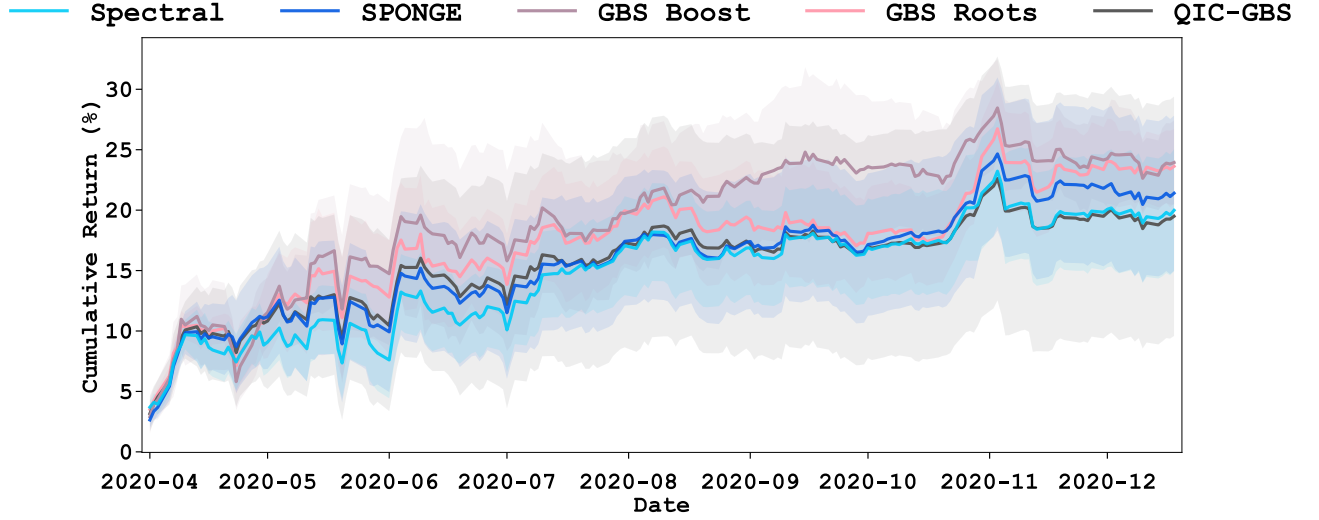}
    }
    \vspace*{-1.2cm} 
    \caption{Cumulative PnL trajectories and standard deviations (coloured bands) for each algorithm. Based on a 60-day beta estimation and a 5-day correlation matrix lookback (Appendix \ref{app:stat_arb_strat}), the effective trading period begins in April 2020. GBS Boost, GBS Roots, and QIC-GBS utilised $0.7N\log N$, $0.6N\log N$, and $N \log N$ samples (See Appendix \ref{app:params}) per GBS call, respectively.}
    \label{fig:100_lossless}
\end{figure*}

We first consider a lossless 100-stock universe ($l_r = 0$). Figure \ref{fig:100_lossless} presents the PnL trajectories. Although all algorithms exhibit similar profiles, quantum methods achieve higher cumulative returns by capturing stronger trades from late May onward, consistent with more effective trade timing and risk management. Although the volatile 2020 sample period introduces stochasticity, the consistently high Sharpe and Sortino ratios for the quantum algorithms (Table \ref{table:lossless_fin}) support the conclusion that the observed gains reflect genuine risk-adjusted performance.

\begin{table}[h]
    \centering
    \renewcommand{\arraystretch}{1.3} 
    \begin{tabular*}{\columnwidth}{@{\extracolsep{\fill}} l r @{$\,\pm\,$} l r @{$\,\pm\,$} l r @{$\,\pm\,$} l @{}}
        \hline
        Method & \multicolumn{2}{c}{TR} & \multicolumn{2}{c}{ShR} & \multicolumn{2}{c}{SoR} \\
        \hline
        Spectral   & 0.200 & 0.051 & 2.035 & 0.512 & 3.389  & 0.931 \\
        SPONGE     & 0.214 & 0.064 & 2.242 & 0.669 & 3.737  & 1.282 \\
        \hline
        GBS Boost  & 0.239 & 0.035 & 2.050 & 0.190 & 3.188  & 0.521 \\
        GBS Roots  & 0.236 & 0.031 & 2.248 & 0.293 & 3.708  & 0.673 \\
        \hline
        QIC-GBS    & 0.195 & 0.099 & 1.781 & 0.893 & 3.089  & 1.696 \\
        \hline
    \end{tabular*}
    \vspace*{-0.5cm}
    \caption{Average total return (TR), Sharpe ratio (ShR), and Sortino ratio (SoR) for each clustering algorithm, with standard deviations across $\sim$20 runs for quantum methods and $\sim$100 runs for classical methods.}
    \label{table:lossless_fin}
\end{table}

As shown in Table \ref{table:lossless_fin}, all algorithms yield competitive financial metrics, but the quantum algorithms deliver the highest returns. Overall, the methods across the financial metrics rank as follows: GBS Roots, SPONGE, GBS Boost, Spectral clustering, and QIC-GBS. Notably, the quantum methods exhibit smaller standard deviations than their classical counterparts, indicating more stable returns in large stock universes, in contrast to the small-universe scaling cases discussed in Appendix \ref{app:mos}.

Despite their financial stability, the quantum methods display lower Jaccard similarities than classical methods (Table \ref{table:lossless_clus}). This disparity suggests that classical algorithms generally produce stable clusters dominated by primary market modes, whereas quantum methods generate more dynamic, structurally varied clusters across windows, which may facilitate high-alpha opportunities. 

GBS Boost yields the highest cluster value, primarily reflecting larger cluster sizes with more contributing correlations. Since our strategy seeks to maximise intra-cluster positive correlations for mean reversion, GBS Boost’s superior average return and cluster value are consistent with this objective. However, GBS Roots achieves comparable returns despite a significantly lower cluster value, suggesting that naively maximising graph features like edge weights does not guarantee optimal economic performance.

\begin{table}[h]
    \centering
    \renewcommand{\arraystretch}{1.3} 
    \begin{tabular*}{\columnwidth}{@{\extracolsep{\fill}} l r @{$\,\pm\,$} l r @{$\,\pm\,$} l r @{}}
        \hline
        Method & \multicolumn{2}{c}{WD} & \multicolumn{2}{c}{J} & \multicolumn{1}{c}{V} \\
        \hline
        Spectral   & 0.111 & 0.023 & 0.344 & 0.055 & 2.756 \\
        SPONGE     & 0.336 & 0.015 & 0.321 & 0.042 & 2.752 \\
        \hline
        GBS Boost  & 0.176 & 0.129 & 0.164 & 0.029 & 17.031 \\
        GBS Roots  & 0.199 & 0.118 & 0.163 & 0.024 & 3.945 \\
        \hline
        QIC-GBS    & 0.262 & 0.173 & 0.154 & 0.025 & 8.151 \\
        \hline
    \end{tabular*}
    \vspace*{-0.5cm}
    \caption{Average weighted density (WD), Jaccard similarity (J), and cluster value (V) of the best partition (by total return) for each algorithm.}
    \label{table:lossless_clus}
\end{table}



This reasoning is further supported by SPONGE achieving the highest WD despite trailing the quantum methods in total returns. SPONGE is intrinsically designed to maximise positive intra-cluster and negative inter-cluster correlations, making it well-suited for this metric. In contrast, GBS Boost optimises for graph density, which may diminish GBS' intrinsic advantage when applied to nearly fully connected graphs. Furthermore, while GBS Roots builds its union using an unordered queue of samples based on weighted densities, this greedy heuristic may be suboptimal; replacing it with a local search procedure could yield superior cluster unions.

SPONGE's low cluster value but high WD indicates that it produces small clusters heavily concentrated with positive correlations. This is expected, since it was designed to maximise positive intra-cluster correlations and negative inter-cluster correlations. While quantum methods exhibit similar behaviour, their larger WD standard deviations may reflect a higher proportion of negative correlations within their clusters.

To validate the performance observed in Figure \ref{fig:100_lossless}, we conducted a two-tailed Welch's t-test comparing the terminal returns of GBS Roots and SPONGE. GBS Roots exhibits a statistically significant advantage in total return at 5\% ($p = 0.023$). Meanwhile, their mean Sharpe ratios are statistically equivalent ($p = 0.949$). However, this parity masks a structural benefit: GBS Roots maintains this competitive risk-adjusted profile with less than half the cross-run standard deviation of SPONGE ($\pm 0.293$ vs. $\pm 0.669$). Thus, the quantum methods consistently deliver stronger and steadier returns.

\subsection{Medium Stock Universe with Loss and Displacement} \label{sec:loss}

\begin{figure*}[t!]
\includesvg[width=\linewidth,height=\textheight,keepaspectratio]{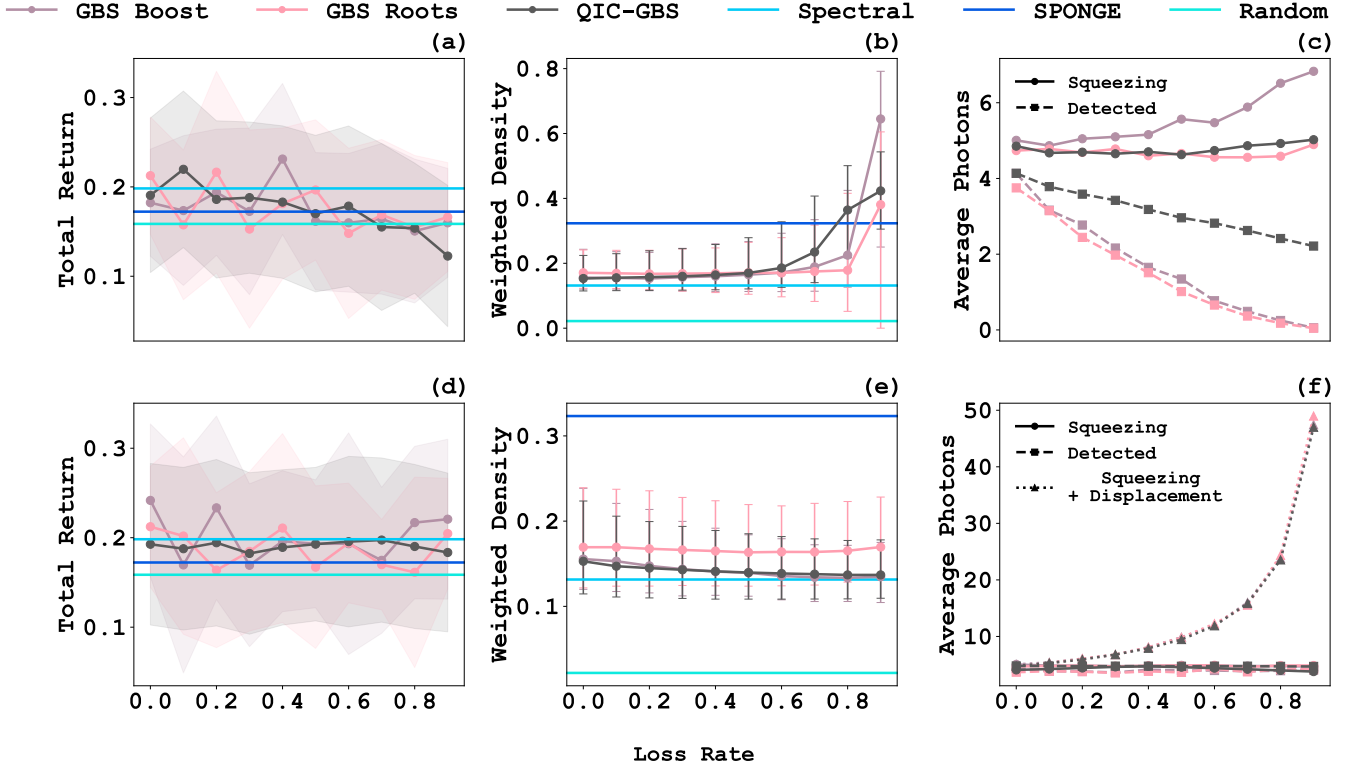}
\vspace*{-0.8cm}
\caption{Performance of GBS-inspired clustering methods under photon loss, with and without displacement compensation. Quantum methods underwent 50 runs per loss rate ($H = 0.7N \log N$ samples per GBS call); classical methods underwent 100-200 runs. The stock universe is $N=50$. The top row (a-c) displays uncompensated loss results, while the bottom row (d-f) incorporates displacement compensation. Classical benchmarks are shown as solid horizontal lines. Panels (a) and (d) show average total returns with standard deviations (coloured bands). Panels (b) and (e) display median weighted densities and interquartile ranges, accounting for left-skewed data. Panels (c) and (f) depict average photon counts during clustering: squeezing only (solid circles), squeezing with displacement (dashed squares), and squeezing with displacement under loss and stochastic effects (dotted triangles).}
\label{fig:loss_p}
\end{figure*}

We evaluate the impact of photon loss ($l_r \in [0, 0.9]$) on GBS clustering within a 50-stock universe. Increasing loss reduces the average photon number; typically this is compensated by increasing squeezing, but doing so distorts the encoded correlation graph and destroys quantum correlations, yielding predominantly classical photons \cite{oh2024classical}. Consequently, under high loss, the sampled subgraphs increasingly reflect a thermal random distribution rather than a true GBS distribution. To isolate this effect, we introduce a random clustering baseline.

Figures \ref{fig:loss_p}(a, b, d, e) show that this random baseline generally underperforms, only exceeding GBS methods at high loss regimes without displacement. Incorporating displacement (see Appendix \ref{app:d_gbs} for details) ensures GBS methods outperform the random baseline across all loss rates. Compared to classical benchmarks, GBS algorithms without displacement lag behind Spectral and SPONGE at loss rates above 70\%. However, displacement stabilises these returns: QIC-GBS consistently outperforms SPONGE (though trailing Spectral), while GBS Boost and GBS Roots frequently outperform both classical methods. Notably, at loss $\le 40\%$, GBS methods beat SPONGE in 85\% of cases. Displacement effectively flattens the performance decay, maintaining robust average returns even in loss-heavy regimes.

The competitiveness of these mixed GBS-thermal distributions at low loss, contrasted with the poor random baseline, implies that pure stochasticity offers no economic edge. Furthermore, these dynamics suggest a decoupling between cluster WD and total return: without displacement, the WD of GBS clusters sharply spikes beyond 60\% loss, even as total returns deteriorate.

This WD spike at high loss is primarily an artifact of low average photon counts. Uncompensated loss yields sparse detections, heavily skewing the sampler toward two-node clusters. Because the union of a maximally weighted pair with additional nodes almost inevitably dilutes the average edge weight, the sampling degenerates into selecting isolated, highly correlated pairs. Thus, the apparent increase in WD from Figure \ref{fig:loss_p}(b) reflects the trivial selection of two-stock clusters rather than robust algorithmic subgraph discovery.

Conversely, applying displacement flattens the WD trajectories in the same manner it flattens returns. Under displacement, GBS methods and Spectral clustering maintain moderate WD, trailing SPONGE, which is explicitly optimised to maximise intra-cluster positive correlations. Ultimately, the divergence between WD and returns under loss, both with and without displacement, demonstrates that strictly maximising positive intra-cluster correlations is insufficient to guarantee optimal economic performance in StatArb portfolios.
Figures \ref{fig:loss_p}(c) and \ref{fig:loss_p}(f) illustrate the impact of photon loss. While input photons from squeezing remain consistent across GBS methods, detected photon counts decay as they traverse the lossy circuit. Because QIC-GBS generates photon pairs via a Poisson distribution, its curve asymptotes to two under high loss. Conversely, GBS Boost and GBS Roots lack this constraint, causing their average photon counts to approach zero. We average these photon counts over all cluster selections. Since we set $\bar{n} = \sqrt{\dim(\mathcal{A})}$, extracting a cluster shrinks the correlation graph, thereby reducing $\dim(\mathcal{A})$ at each step. Thus, the total average photon count reflects the average cluster size generated by the GBS algorithms. 

In particular, Figure \ref{fig:loss_p}(f) demonstrates how displacement affects these average photon counts. Compensating for photon loss by injecting coherent light via displacement dramatically increases input photons at high loss regimes (dotted curve). Consequently, detected photon counts remain relatively stable across all loss rates, trailing only slightly behind the original squeezed input photons due to loss-induced stochastic effects.
\vspace*{\fill}
\subsection{Medium Stock Universe without Loss and Across Different Market Regimes}
We evaluate algorithm performance across four distinct market regimes: 2008 with its high-volatility bear market, 2017 with its low-volatility bull market (where StatArb typically underperforms due to limited mean-reversion opportunities), 2020 featuring a highly dynamic hybrid market with a sharp initial crash, rapid recovery, and significant sector rotations, and 2022 with its prolonged hybrid-bear market). This selection provides a rigorous test of the clustering algorithms' agility in adapting to sudden regime shifts versus extended macroeconomic drawdowns. 

\begin{figure}[t!]
\includegraphics[width=\linewidth,height=\textheight,keepaspectratio]{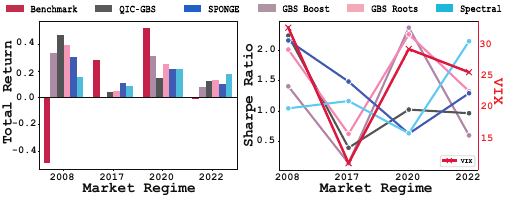}
\caption{Clustering method performance across disparate market regimes using a 50-stock universe. (Left) Total return of all clustering methods alongside a benchmark for 2008, 2017, 2020, and 2022. (Right) Comparison of algorithm Sharpe ratios against the average S\&P 500 Volatility Index (VIX) per regime. Quantum methods utilised $H = 0.7N\log N$ samples per GBS call.}
\label{fig:diff_year}
\end{figure}

The left plot of Figure \ref{fig:diff_year} benchmarks the clustering methods against a static $1/N$ equal-weighted buy-and-hold benchmark. During the 2008 crash, the GBS methods particularly generate robust positive returns. This highlights the StatArb strategy's capacity to isolate idiosyncratic alpha and maintain market neutrality during broad-market sell-offs. Conversely, in the low-volatility 2017 regime, GBS methods underperform.

In the highly dynamic 2020 regime, quantum algorithms generate the highest returns, demonstrating their adaptability to rapid structural changes. However, their edge diminishes during the 2022 bear market, albeit they remain competitive. This suggests that quantum algorithms excel at capturing rapid mean-reverting dislocations, but may struggle to maintain dominance during extended, unidirectional trends.

The right plot of Figure \ref{fig:diff_year} reveals a strong positive correlation between the Sharpe ratios of the quantum methods and the VIX. Their ability to deliver high risk-adjusted returns during the extreme volatility of 2008 and 2020 is consistent with  stronger market neutrality compared to classical methods. 


\section{Discussion} \label{sec:disc}

A key assumption, for simplicity, in our simulations is that the trades executed do not incur price impact and there are zero transaction costs.

Although we frame portfolio clustering as a MACCP \cite{bansal2004correlation}, our empirical results across varying stock universes and loss rates indicate that strictly maximising graph-based features (such as WD or value) does not guarantee optimal economic performance. While poorly performing methods consistently exhibit weak WD, the highest-returning methods do not necessarily exhibit the strongest WD, highlighting that WD is a useful but incomplete proxy for economic performance. This may occur because if the correlation between two assets is too strong, there may be fewer price deviation opportunities for mean reversion.

Under low-loss regimes, quantum methods sample from a mixed distribution, suggesting that physical photon loss acts as an implicit regulariser against overfitting to the absolute densest subgraphs. See Appendix \ref{app:lossy_gbs} for more details on lossy GBS. Introducing displacement~\cite{Thekkadath2022ExperimentalDisplacement} when loss is present  further mixes the DGBS/GBS distribution with a thermal distribution, injecting stochasticity into cluster determination (analogous to findings in \cite{liu2025stochastic}). Consequently, traversing a broader, noise-injected solution space can provide  beneficial cluster diversification compared to pure deterministic graph optimisation.




Furthermore, current GBS methods are not natively optimised for the MACCP. GBS Boost evaluates candidate clusters using graph density, ignoring correlation. Because financial correlation graphs are nearly fully connected, extracting maximum-density clusters is trivial. Pre-processing the adjacency matrix to drop negative correlations induces sparsity, thereby restoring GBS' intrinsic advantage in dense subgraph identification. 

To perform better in stable market regimes, the algorithm should be designed to recognise and retain profitable partitions throughout the sample period. Ultimately, GBS is a probabilistic sampling heuristic; relying on it for combinatorial optimisation without a tailored classical subroutine is limiting. Therefore, future work should explore hybridising GBS with rigorous combinatorial solvers to further amplify StatArb returns.

An alternative cluster evaluation metric could leverage the Ornstein-Uhlenbeck process to quantify the strength of alpha signals \cite{uhlenbeck1930theory, maller2009ornstein, elliott2005pairs, krauss2017statistical}. By modelling spread dynamics as a stochastic differential equation, a linear autoregressive model can estimate the mean-reversion speed $\theta$ for each cluster, thereby prioritising subgraphs with faster trading opportunities. Notably, assuming uniform latent factors and idiosyncratic noise within a cluster, the optimal WD to maximise expected returns is $2/3$. Although this setting is idealised, this analytically underscores that maximising intra-cluster correlation does not monotonically increase economic returns.

Regarding QIC-GBS, the algorithm produced the lowest risk-adjusted returns and highest variance, yet yielded the second-highest WD and cluster value. This suggests a propensity for assembling large, highly positively correlated clusters that paradoxically fail to yield profitable arbitrage spreads. Under loss, its returns decayed monotonically but stabilised when compensated by displacement. Across market regimes, its performance was highly idiosyncratic: it dominated the 2008 crash while lagged in the 2020 hybrid market. Nevertheless, owing to its classical formulation and rapid simulation times, QIC-GBS remains a pragmatic proxy for GBS dynamics.

Although our one-year periods demonstrated the quantum methods' greater market neutrality during volatile periods, their economic edge degraded during stable regimes (evidenced by lower inter-window Jaccard similarities). Future research should evaluate multi-year periods like the 2007–2009 financial crisis, 2017–2018 speculative bubbles, or longer, to assess algorithmic adaptability across prolonged macroeconomic regime shifts. 


\section{Conclusion} \label{sec:conc}

This work presents a novel application of GBS to financial trading, specifically for clustering assets within a StatArb strategy~\cite{jin2023correlation}. By interpreting residual correlation matrices as signed weighted graphs and pre-processing them into non-negative adjacency matrices suitable for a GBS device, we introduced and evaluated two quantum clustering algorithms: GBS Boost~\cite{bonaldi2024boost} and GBS Roots. These methods leverage the intrinsic bias of GBS toward dense subgraphs and a clustering subroutine, to identify robust clusters of co-moving assets. This framework successfully generated mean-reverting trading signals, ultimately yielding higher returns in a volatile market for a lossless 100-stock universe.

Our study also identifies clear boundaries regarding physical implementation. Under photon loss rates exceeding 60\%, the performance of GBS methods degraded below that of classical competitors, approaching a random clustering baseline. However, introducing displacement operations~\cite{mer2026} to compensate for this loss (aiming to preserve the average photon number across varying loss rates) effectively stabilised both returns and weighted densities. Consequently, displacement enables quantum methods to remain economically competitive with classical algorithms at high loss levels, decisively outperforming random baselines.

Ultimately, this work establishes the potential of GBS for trading portfolio optimisation. Future research should prioritise implementing these algorithms on physical photonic hardware, though demonstrating definitive quantum advantage will require devices with a large number of modes. Additionally, integrating advanced signal processing techniques, such as Ornstein-Uhlenbeck parameter estimation, could further refine cluster selection. Despite current hardware constraints, quantum clustering provides a promising, mathematically distinct heuristic for identifying economic market structures, particularly when applied to sufficiently large and volatile market subsets under low-loss conditions.

\section*{Data \& code availability}
The data and code used in the simulations are available at \cite{dml_doi}.
\vspace*{\fill}
\begin{acknowledgments}
This work was supported by UK Research and Innovation Future Leaders Fellowship (project: MR/W011794/1) and Guarantee Postdoctoral Fellowship (project: EP/Y029631/1), Engineering and Physical Sciences Research Council (EPSRC) UK Quantum Technologies Program's hub for Quantum Computing via Integrated and Interconnected Implementations (project: EP/Z53318X/1). D.M.L. acknowledges funding support from QuEST (project: EP/W524323/1).
\vspace*{\fill}
\end{acknowledgments}

\section*{Disclaimers}
This paper was prepared for information purposes and is not a product of HSBC Bank Plc. or its affiliates. Neither HSBC Bank Plc. nor any of its affiliates make any explicit or implied representation or warranty and none of them accept any liability in connection with this paper, including, but not limited to, the completeness, accuracy, reliability of information contained herein and the potential legal, compliance, tax or accounting effects thereof.
\bibliography{ref}
\appendix

\section{Statistical arbitrage strategy} \label{app:stat_arb_strat}

Our empirical investigation begins with daily close price and dividend data sourced from Wharton Research Data Services \cite{wrds}. Let $P_{t,i}$ and $D_{t,i}$ represent the price and dividend of stock $i$ at time $t$, yielding the total return $R_{t,i} \;=\; \frac{P_{t,i}+ D_{t,i}}{P_{t-1,i}} - 1$. To isolate the idiosyncratic dynamics of each stock, we strip away broader market exposure by calculating the market residual returns (MRRs). The MRR is defined as $R_{t,i}^{\text{res}} := R_{t,i} -\beta_i R_{\text{mkt},t}$, where $\beta_i := \frac{\text{Cov}(R_i, R_{\text{mkt}})}{\text{Var}(R_{\text{mkt}})}$ quantifies the stock's sensitivity to market movements. Neutralising these systemic factors ensures that extracted alpha is generated primarily by clustering  rather than underlying market beta.

These residual returns form the basis of an $N \times N$ correlation matrix $\mathcal{C}$ at time $T$, functioning as the adjacency matrix of a weighted, signed graph. We compute the entries over a $w$-day rolling window:

\begin{align} \label{cmrr}
    \mathbf{\mathcal{C}}_{ij} \;:=\; 
        \frac{
        \sum_{t=T-w}^{T-1} 
        \bigl(R^{\text{res}}_{t,i} - \bar{R}^{\text{res}}_{i}\bigr)
        \bigl(R^{\text{res}}_{t,j} - \bar{R}^{\text{res}}_{j}\bigr)
        }{
        (w-1)\,\sigma_i \, \sigma_j
        }.
\end{align}

\noindent As the window rolls forward and prices update, this underlying FCG dynamically evolves. 

\begin{figure*}[ht!]
\includegraphics[width=\linewidth, trim=0cm 0cm 0cm 4cm, clip]{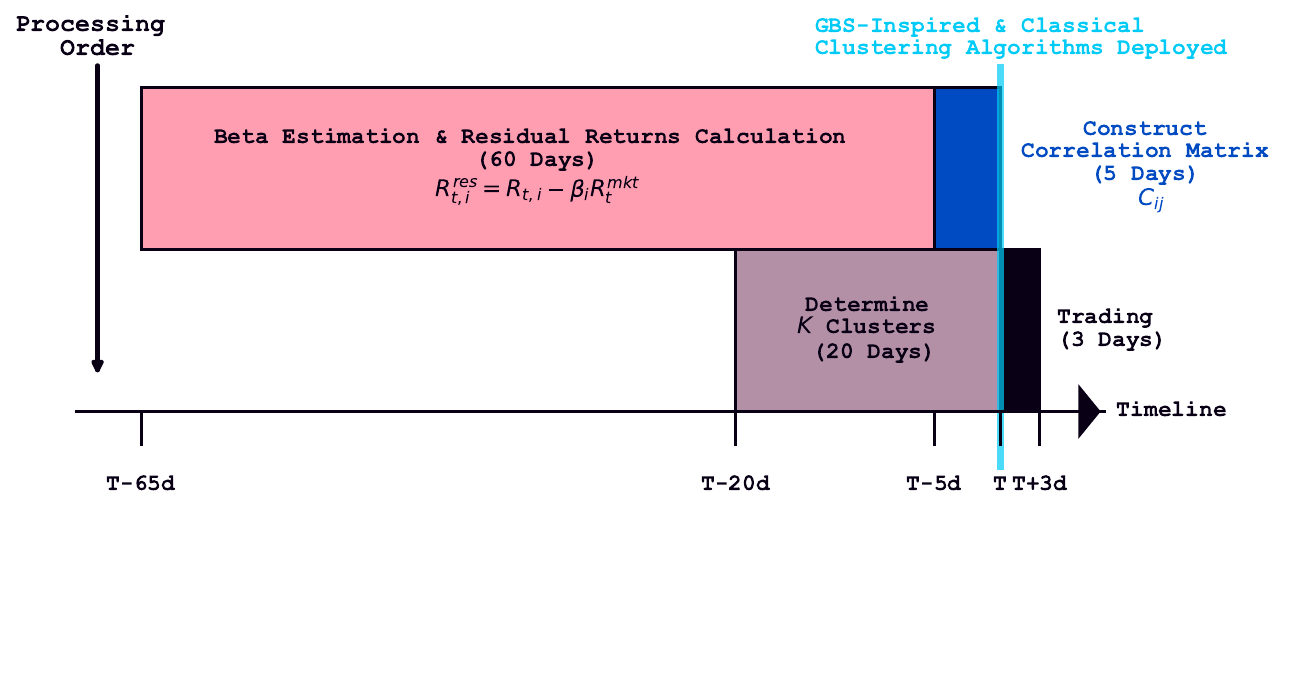}
\vspace*{-3cm}
\caption{A snapshot of the StatArb strategy timeline. The vertical axis indicates the processing hierarchy, while the horizontal axis denotes chronological execution. We sequentially compute beta values, residual returns, and the residual correlation matrix. For classical methods, this matrix dictates the cluster count per rolling window. Clustering algorithms are deployed at day zero (time T) for subsequent trading. This framework rolls forward every three days (unless triggered earlier by the stop-win threshold) across a 252-day trading year.}
\label{fig:timeline}
\end{figure*}

As illustrated in the strategy timeline in Figure \ref{fig:timeline}, we utilise a 60-day window to estimate betas, followed by a successive 5-day lookback period to construct the correlation matrix. Trailing this is a 20-day cluster determination period, aligned to the end of the 5-day window, where the target cluster count is calculated for the classical methods.

To ensure the clustering algorithms operate on genuine correlation structures, we filter $\mathcal{C}$ using Random Matrix Theory (RMT) or explained variance approaches \cite{jin2023correlation}. According to the Marchenko-Pastur (MP) distribution, eigenvalues falling within the bulk $[\lambda_-,\lambda_+]$ represent random noise. We construct a cleaned correlation matrix by nullifying the market mode (the largest eigenvector) and the noise band (eigenvalues within the MP bulk), projecting the matrix strictly onto its statistically significant principal components corresponding to eigenvalues greater than $\lambda_+$. See Appendix \ref{app:eval_spacings} for definitions and the full extension of the RMT approach.

For classical clustering pipelines, the target cluster count $K$ is determined dynamically by the number of these dominant eigenvalues. Conversely, our quantum methods do not strictly enforce a dynamic target cluster count. This design choice allows the partitions to naturally reflect GBS sampling dynamics, avoiding forced truncation steps that could destroy GBS-derived clusters. 

One could argue if the reason quantum methods performed stronger than classical methods is because they are unconstrained by $K$. However, from Appendix \ref{app:eval_spacings}, we will see that $K$ dynamically changes due to the market. Therefore, these classical methods are intrinsically equipped with sensitivity of the prevailing macroeconomic regime. Thus, they are guided by a market-informed `oracle' that dictates the exact number of structural factors present in three day intervals.

Conversely, the quantum methods operate entirely blind to this macro-parameter, partitioning the graph organically based solely on topological subgraph sampling. This deliberate methodological asymmetry ensures that any outperformance by the GBS methods can be strictly attributed to their intrinsic clustering logic, rather than an artificially optimised cluster count.

All clustering algorithms are deployed immediately after the end of the 5-day correlation period. Following cluster assignment, we extract arbitrage signals based on cross-sectional mean reversion. Within a specific cluster $j$, we calculate the cross-sectional mean return $\bar{R}_{t,j} = \frac{1}{j_n} \sum_{i=1}^{j_n} R_{t,i}$. A stock $j_i$ is classified as a Previous Winner (PW) or Previous Loser (PL) based on its cumulative deviation $\Delta_{t,j_i} = R_{t,j_i} - \bar{R}_{t,j}$ over the $w$-day formation window. Utilising a threshold $p \ge 0$ (set to $p=0$ in this strategy), the classification rule is:

\begin{align}
j_i = \begin{cases}
    \text{PW}, \quad \text{if} \quad \sum_{t = T-w}^{T-1} \Delta_{t,j_i} > p, \\
    \text{PL}, \quad \text{if} \quad \sum_{t = T-w}^{T-1} \Delta_{t,j_i} < -p,
\end{cases}
\end{align}

\noindent The magnitude of the trading signal $\Delta_{t,j_i}$ acts as a confidence proxy; extreme deviations imply a higher probability of impending mean reversion toward the cluster mean.

We deploy a zero-cost, dollar-neutral portfolio constrained by a strict capital budget. Capital is assigned uniformly across the identified PW and PL constituent legs. To achieve market-neutrality, the portfolio normalises weights such that the aggregate absolute dollar value of the short positions (PWs) equals the long positions (PLs), summing to unit capital per cluster. For example, if a cluster contains 8 winners and 2 losers, we allocate \$0.125 to each short position and \$0.5 to each long position. 

Returns are generated by exploiting short-term price deviations under the anticipation of mean reversion. By short-selling PWs and taking long positions in PLs at time $t$, our total return for a cluster $M$ over holding period $l$ (with $N$ stocks comprising $n$ PLs $i_1,\hdots,i_n$ and $m$ PWs $j_1, \hdots, j_m$) is:

\begin{align}
    r(t,l,M) = \sum_{a =1}^n \left(p_{t+l,i_a} - p_{t,i_a}\right) + \sum_{b =1}^m \left(p_{t, j_b} - p_{t+l, j_b}\right).
\end{align}

\noindent To elaborate, by short-selling PWs, we borrow shares of a stock $j$ then immediately sell it at time $t$ at price $p_{t, j}$, we expect the price of stock $j$ to decrease to $p_{t+l, j}$ after $l$ days, then we have to return the borrowed shares of stock $j$ so that our profit is $p_{t, j} - p_{t+l, j}$. On the other hand, when we hold long positions in PLs, we buy a stock $i$ at time $t$ at price $p_{t, i}$, after $l$ days we expect the price of the stock to increase to $p_{t+l, i}$, we then sell stock $i$ such that our profit is $p_{t+l,i} - p_{t,i}$.

At the end of the $l=3$ day holding period, the portfolio is liquidated, returns are recorded, and a re-balancing process shifts the entire operational window forward by 3 days. This requires the re-calculation of betas, correlations, target clusters, and signals. To mitigate downside risk and capture rapid mean-reverting transients, we overlay a stop-win threshold $q=0.02$. If the cumulative portfolio return breaches $q$ prior to the end of the 3-day period, this indicates successful mean reversion, triggering immediate early liquidation and subsequent portfolio re-balancing.

\section{Cluster determination from eigenvalue spacings} \label{app:eval_spacings}

We introduce a modified method for determining the optimal number of clusters required by classical clustering algorithms within our StatArb framework. This approach addresses the limitations of traditional dominant eigenvalue and explained variance methods, which systematically underestimated and overestimated the number of clusters, respectively, for our specific stock universe and sample period. These estimation errors previously caused classical algorithms to group the majority of stocks into a single cluster or fail to meet the target cluster count entirely.

Recalling the correlation matrix of residual returns from Equation (\ref{cmrr}), consider $N$ stocks over a sample period of $T$ days. The residual returns can be stored in a $T \times N$ matrix $\mathbf{X}$ defined by:

\begin{align}
    X_{t,i} := \frac{R^{\text{res}}_{t,i}-\bar{R}^{\text{res}}_{i}}{\sigma_i}.
\end{align}

\noindent This allows us to rewrite Equation (\ref{cmrr}) as $\mathbf{\mathcal{C}} = \frac{1}{w-1}\mathbf{X}^{\mathsf{T}}\mathbf{X}$. By construction, each column of $\mathbf{X}$ possesses a sample mean of 0 and a sample variance of 1. To test for the presence of a non-random correlation structure, we compare the empirical matrix $\mathbf{\mathcal{C}}$ against a theoretical null hypothesis, $H_0$, which assumes that the residual returns consist of pure, uncorrelated noise.

Under $H_0$, the entries $X_{t,i}$ are independent and identically distributed (i.i.d.) random variables with a population mean of 0 and a variance of 1. Assuming $X_{t,i} \sim \mathcal{N}(0,1)$, the theoretical matrix $\mathbf{\mathcal{C}}$ follows a Wishart ensemble, $\mathcal{W}_N(I, T)$, serving as our pure noise benchmark. Let $\rho := \frac{N}{T}$ denote the ratio of stocks to the sample period length. Under the null hypothesis, for a fixed $\rho$ as $N, T \to \infty$, the Marchenko-Pastur (MP) theorem states that the empirical limiting spectral distribution of $\mathbf{\mathcal{C}}$ converges to the MP distribution. This distribution is defined by the density function:

\begin{align}
    p_{MP}(\lambda) = \begin{cases}
        \frac{\sqrt{(\lambda_+-\lambda)(\lambda_--\lambda)}}{2 \pi \lambda \rho}, \quad &\lambda \in [\lambda_-, \lambda_+] \\
        0, \quad &\text{otherwise}
    \end{cases},
\end{align}

\noindent where $\lambda_{\pm} = (1 \pm \sqrt{\rho})^2$. Consistent with the methodology in~\cite{jin2023correlation}, we are interested in eigenvalues $\lambda$ of $\mathbf{\mathcal{C}}$ that strictly exceed the upper edge of the bulk, $\lambda_+$. The number of eigenvalues greater than $\lambda_+$ determines the number of clusters in a window, as these significant eigenvalues correspond to dominant structural signals in the stock returns. However, in our StatArb strategy, $T$ is fixed while the stock universe varies. As $N$ increases, $\rho$ and consequently $\lambda_+$ also increase, creating a steep upper bound that empirically yields zero to two clusters for $K$.

To overcome this limitation, we adopt a cluster determination method grounded in the spiked covariance model~\cite{johnstone2001distribution, baik2006eigenvalues}. For a clustered system, the population covariance matrix $\mathbf{\Omega}$ is given by:

\begin{align} \label{scm}
    \mathbf{\Omega} = \mathbf{I}_N + \sum_{k=1}^K \theta_k v_k v_k^{\mathsf{T}},
\end{align}

\noindent where the spike strengths $\theta_k \geq 0$ indicate underlying factors causing structural co-movement above the noise, and the orthonormal vectors $v_k$ define the direction of these spikes. This perturbation of the identity matrix yields eigenvalues 

\begin{align}
    \nu_k = \{1+\theta_k\}_{k=1}^K \cup \underbrace{\{1,\hdots,1\}}_{N-K \; \text{times}}.
\end{align}

We relate this model to our empirical correlation matrix via the transformation $\mathbf{X = Z\Omega^{\frac{1}{2}}}$, where $Z_{t,i} \sim \mathcal{N}(0,1)$. Under $H_0$, no spikes exist ($\mathbf{\Omega} = \mathbf{I}_N$ and $\mathbf{X} = \mathbf{Z}$), returning the original Wishart ensemble whose eigenvalues form the MP bulk. Conversely, when spikes occur, we focus specifically on supercritical spikes ($\theta_k > \sqrt{\rho}$), as these correspond to $\nu_k$ detaching from the bulk.

For each supercritical spike, the corresponding sample eigenvalue of $\mathbf{\mathcal{C}}$ detaches from the bulk and converges to the limit dictated by the BBP phase transition~\cite{baik2005phase}: $\lambda_k = (1+\theta_k)(1+\frac{\rho}{\theta_k})$. To distinguish whether the largest observed eigenvalues represent genuine BBP signals or extreme noise fluctuations from the MP bulk, we must examine the distribution of the largest noise eigenvalue, $\lambda_1$, under the null hypothesis. The fluctuations of this upper bulk edge follow the Tracy-Widom (TW) distribution. We define the test statistic~\cite{johnstone2001distribution}:

\begin{widetext}
\begin{align}
    z_1 := \frac{\lambda_1 - \mu_{N,T}}{\sigma_{N,T}} 
    \; \xrightarrow\; \mathrm{TW}_1 \sim F_1(y) = \text{exp}\left\{-\frac{1}{2} \int_y^{\infty} \bigl(q(x) + (x-y)q^2(x)\bigr) dx\right\},
\end{align}
\end{widetext}

\noindent where $\mu_{N,T} = \frac{1}{T-1}(\sqrt{N} + \sqrt{T-1})^2$ defines the location of the upper bulk edge $\lambda_+$, and $\sigma_{N,T} = \frac{1}{T-1}(\sqrt{N} + \sqrt{T-1}) \bigl(\frac{1}{\sqrt{N}} + \frac{1}{\sqrt{T-1}}\bigr)^{\frac{1}{3}}$ scales the fluctuations of the top eigenvalue around $\lambda_+$. Asymptotically, $z_1$ converges to a first-order TW distribution, where $q(x)$ behaves asymptotically as the Airy function $\text{Ai}(x)$ as $x \to \infty$.

Following~\cite{johnstone2001distribution}, we can conduct a hypothesis test to validate the spiked covariance model: for a significance level $\alpha$, if $z_1$ exceeds the $(1-\alpha)$-quantile of $\mathrm{TW}_1$, we reject the null hypothesis and accept $\lambda_1$ as a supercritical spike with $(1-\alpha)$ confidence. As $T \to \infty$, the scaling factor simplifies to $\sigma_{N,T} = \sqrt{\rho}(1+\sqrt{\rho})^{\frac{4}{3}}N^{-\frac{2}{3}}$, confirming that for a fixed $\rho$, the fluctuation scale is $\mathcal{O}(N^{-\frac{2}{3}})$.

We utilise eigenvalue spacings to determine the optimal cluster count. For the ordered eigenvalues of $\mathbf{\mathcal{C}}$, the $k$th spacing is defined as $s_k := \lambda_k - \lambda_{k+1}$. We employ spacings because the distribution for real Wishart matrices follows $p_W(s) \sim a s e^{-b s^2}$ ($a,b \in \mathbb{R}$), demonstrating exponential repulsion. Consequently, large spacings are exceedingly rare under pure noise and strongly indicate the presence of a spike.

Under $H_0$, spacings at the upper bulk edge also scale as $s_k \sim \mathcal{O}(N^{-2/3})$ according to the Airy point process, which extends the TW distribution to higher-order top eigenvalues ($\lambda_1, \lambda_2, \lambda_3$, etc.). We define a spacing as significant if $s_k > \tau$, where the threshold is $\tau = q N^{-\frac{2}{3}}$, and $q$ is a quantile correction constant dynamically recalculated for finite $N$ and $T$ in each window.

To prevent an arbitrary threshold from yielding inaccurate cluster counts, we empirically calibrate $\tau$. We generate $B$ null replicates, $\mathbf{X}_B$, via column-wise permutations of $\mathbf{X}$. This permutation preserves the marginal statistical properties of the data while destroying the underlying correlation structure, effectively simulating pure noise. We compute the top spacings $s_{1,B}$ across these replicates to build a null distribution. Extracting the $(1-\alpha)$-quantile, $q_{1-\alpha}$, provides our critical threshold: $\tau = q_{1-\alpha}N^{-\frac{2}{3}}$.

We formalise cluster determination through a sequence of right-tailed hypothesis tests:
\begin{itemize}
    \item \textbf{Hypotheses:} $H_0^{(k)}: s_k \leq \tau$, \; $H_1^{(k)}: s_k > \tau$, for $k=1,2,\hdots,K$.
    \item \textbf{Test Statistic:} $s_1 = \lambda_1 - \lambda_2$.
    \item \textbf{Significance Level:} $\alpha = 0.05$.
    \item \textbf{Decision Rule:} Reject $H_0^{(k)}$ if $s_k > \tau$. A rejection at $k=1$ implies the existence of at least one cluster.
\end{itemize}

Detectable spikes correspond to eigenvalues detaching from the MP bulk, with the TW distribution governing their $\mathcal{O}(N^{-\frac{2}{3}})$ scale. This sequential testing terminates at the first failure to reject the null hypothesis. The estimated cluster count is therefore defined as:

\begin{align}
    \hat{K} := \max\{k \ge 1 : s_1 > \tau, \hdots, s_k > \tau\},
\end{align}

\noindent where $\hat{K}=0$ if $s_1 \leq \tau$. This sequential termination controls the family-wise error rate (FWER), which is the probability of making type I errors across multiple tests. Because the FWER is bounded by $\alpha$, we can claim with $(1-\alpha)$ confidence the existence of $\hat{K}$ clusters. Under the joint null hypothesis, the top spacings governed by the Airy point process are asymptotically independent at the $\mathcal{O}(N^{-\frac{2}{3}})$ scale; thus, the probability of $k$ consecutive false rejections is $\alpha^k \leq \alpha$.

While the TW test statistic $z_1$ accurately identifies supercritical spikes asymptotically (as $N, T \to \infty$ and $\rho \to \text{const.}$), our finite setting ($N \in [12, 100]$ and $T=20$) causes the asymptotic TW description to break down. This structural limitation necessitates the permutation-based, dimensionless bootstrapping calibration threshold $\tau$, subject to a calibration error $\varepsilon_B$.

This calibration error $\varepsilon_B$ quantifies the finite-sample limitation of the bootstrap. For the $B$ permuted null replicates of top spacings $s_{1,b}$ (where $b \in \{1,\hdots,B\}$), the spacings are i.i.d. and drawn from the true null distribution $F_1$. Let $\hat{F}_B$ denote their empirical cumulative distribution function. For any $\varepsilon > 0$, the Dvoretzky-Kiefer-Wolfowitz (DKW) inequality~\cite{dvoretzky1956asymptotic, massart1990tight} establishes:

\begin{align}
    P\Bigl(\sup_x |\hat{F}_B(x) - F_1(x)| \le \varepsilon\Bigr) \ge 1 - 2e^{-2B\varepsilon^2}.
\end{align}

\noindent Setting $\varepsilon = \varepsilon_B$ and defining $1-\delta$ as the probability that the error remains bounded by $\varepsilon_B$, we obtain $\varepsilon_B = \sqrt{\frac{\log(2/\delta)}{2B}}$. Because the empirical quantile satisfies $\hat{F}_B(q_{1-\alpha}) = 1-\alpha$, the DKW bound guarantees that $|F_1(q_{1-\alpha}) - (1-\alpha)| \leq \varepsilon_B$ with probability $1-\delta$. For instance, with $B=1000$ permutations and a confidence level of $1-\delta = 0.95$, the maximum calibration error is strictly bounded to $\varepsilon_{1000} \approx 0.043$.

\section{Gaussian boson sampling} \label{app:gbs_theory}
In GBS, Gaussian states are injected into a Haar-random interferometer composed of linear optical elements. Because these elements preserve Gaussianity, the resulting output state is fully characterised by a $2m \times 2m$ covariance matrix $\mathbf{\Sigma}$ and a displacement vector, which we assume to be zero. From this covariance matrix, we derive the matrix $\mathbf{B} = \mathbf{Y}(\mathbf{I} - \tilde{\mathbf{\Sigma}}^{-1})$, where $\mathbf{Y} = \begin{pmatrix} 0 & \mathbf{I}_m \\ \mathbf{I}_m & 0 \end{pmatrix}$, $\tilde{\mathbf{\Sigma}} = \mathbf{\Sigma} + \mathbf{I}/2$, and $\mathbf{I}$ and $\mathbf{I}_m$ are appropriately sized identity matrices. Because $\mathbf{B}$ is symmetric, it can be partitioned into four $m \times m$ block matrices comprising squeezed ($\mathbf{S}$) and thermal ($\mathbf{T}$) contributions~\cite{kruse2019detailed}:

\begin{align} \label{gbs_block_decom}
    \mathbf{B} = \left( \begin{array}{c|c}
    \mathbf{S} & \mathbf{T} \\ 
    \hline
    \mathbf{T}^{\mathsf{T}} & \mathbf{S}^* \\
\end{array}\right).
\end{align}

\noindent Assuming pure Gaussian states implies $\mathbf{T}=0$, which reduces Equation (\ref{gbs_block_decom}) to $\mathbf{B} = \mathbf{T} \oplus \mathbf{T}^*$. The hafnian, denoted as $\text{Haf}(\mathbf{B})$, is a matrix function that calculates the number of ways a set of $2m$ objects can be paired. It dictates the underlying probability distribution from which GBS draws its samples and is defined as:

\begin{align} \label{haf}
    \text{Haf}(\mathbf{B}) := \sum_{\sigma \in \textrm{PMP}} \prod_{(i,j) \in \sigma}B_{ij},
\end{align}

\noindent where PMP denotes the set of perfect matching permutations of $2m$ objects. Detecting photons at the output yields a photon number pattern $\mathbf{n} = (n_1, \dots, n_m)$, where $n_k$ is the number of photons detected in mode $k$. The probability of observing a specific configuration $\mathbf{n}$ is proportional to the squared magnitude of the hafnian of a specific submatrix:

\begin{align} \label{gbs_prob_main}
    \text{Prob}(\mathbf{n}) = \frac{1}{\sqrt{\text{Det}\left(\mathbf{\Tilde{\Sigma}}\right)}}\frac{|\text{Haf}(\mathbf{B}_{\mathbf{n}})|^2}{\displaystyle \prod_{k=1}^m n_k!},
\end{align}

\noindent where $\mathbf{B}_{\mathbf{n}}$ is the submatrix of $\mathbf{B}$ constructed by repeating the $k^{\text{th}}$ row and column $n_k$ times to reflect the detected photons. This construction yields an $N \times N$ square matrix, where the dimension grows based on the total number of detected photons, $N = \sum_{k=1}^m n_k$ (with some abuse of notation).

\section{Pre-processing the input GBS matrix and alternatives} \label{app:pre-pro}

FCGs are constructed from correlation matrices $\mathbf{\mathcal{C}}$ with entries $\mathbf{\mathcal{C}}_{ij} \in [-1,1]$. To model these as undirected, loopless graphs suitable for encoding into a Gaussian Boson Sampling (GBS) device, we zero the main diagonal. Thus, the input adjacency matrix becomes $\mathbf{\mathcal{A}} = \mathbf{\mathcal{C}} - \mathbf{I}$, where $\mathbf{\mathcal{A}}_{ij} \in [-1,1]$ for $i \neq j$.

If the input matrix for the GBS device contains exclusively non-negative entries, classical simulation of the hafnian is generally believed to be feasible in polynomial time~\cite{jerrum2004polynomial, uvarov2024performance, Oh2024quantum-inspired}, potentially nullifying the quantum advantage of GBS. Indeed, \cite{anand2025simulating} recently demonstrated this by employing a Jerrum--Sinclair Markov chain to sample perfect matchings from the Cartesian product graph $\mathcal{G} \Box K_2$, successfully generating vertex subsets distributed proportionally to the squared hafnian in polynomial time.

Despite this risk to classical intractability, non-negative edge weights remain highly desirable for graph clustering applications. They preserve the intuitive relationship where higher edge weights correlate directly with an increased probability of GBS sampling those subgraphs. Furthermore, retaining negative edges introduces a critical structural flaw: sampled subgraphs containing perfect matchings with an even number of negative pairs yield an undesirable positive contribution to the hafnian. Consequently, the GBS device would paradoxically show an increased probability of sampling subgraphs dominated by negative correlations in perfect matching pairs.

To ensure our clustering is strictly guided by positive price co-movement, we apply a thresholding function to the adjacency matrix, $t: \mathbf{\mathcal{A}}_{ij} \mapsto \max(0, \mathbf{\mathcal{A}}_{ij})$, dropping all negative values. While this prevents the sampling of negatively correlated subgraphs, it risks sparsifying the graph and threatening the classical hardness of GBS~\cite{yang2024speeding, oh2022classical}. However, this thresholding crucially restores the utility of graph density; without it, the fully connected nature of raw FCGs would result in all sampled subgraphs exhibiting maximum density.

Fortunately, extreme sparsity is rare in FCGs. Unprocessed FCGs are dominated by positive correlations driven by the global market mode, which is a spectral component representing the largest eigenvalue and acting as a proxy for systemic market risk. In our raw dataset, prior to stripping market factors, $99.53\%$ of correlations are positive. 

Isolating residual returns strips these broad market factors, dropping the proportion of positive correlations to $51.36\%$; however, the resulting graphs remain sufficiently dense. Adopting the threshold for sparsity from~\cite{roughgarden2018algorithms} (where a graph is sparse if the number of edges $|E|$ scales as $\mathcal{O}(|V|\log|V|)$), even our smallest graph ($|V|=12$ vertices) comfortably exceeds this bound after negative values are removed. 

We now evaluate alternative pre-processing methods and detail our rationale for rejecting them. Several alternative transformations can render the input GBS matrix non-negative. For instance, continuous mappings such as $t_1: \mathbf{\mathcal{A}}_{ij} \mapsto \exp(\mathbf{\mathcal{A}}_{ij})$ or $t_2: \mathbf{\mathcal{A}}_{ij} \mapsto \frac{\mathbf{\mathcal{A}}_{ij} + 1}{2}$ preserve the information from negative correlations by compressing them into small, pseudo-positive values. Similarly, positive correlations are scaled to larger values, ensuring that sampled subgraphs remain predominantly composed of positively correlated edges.

However, these transformations lack a hard noise floor. Correlations near zero are mapped to values near $1$ (for $t_1$) or $0.5$ (for $t_2$). This artificial inflation can bloat the sampling probability of clusters dominated by random noise. Furthermore, while $t_1$ and $t_2$ suppress the influence of negative correlations, yielding lower hafnians than purely positive graphs, their probability of being sampled remains non-negligible. Consequently, undesirable, negatively correlated subgraphs can still be generated. In contrast, our chosen thresholding function $t$ eliminates these undesirable graphs, guaranteeing that the portfolio clustering is driven entirely by positive co-movement.

Alternatively, applying an absolute value transformation, $t_3: \mathbf{\mathcal{A}}_{ij} \mapsto |\mathbf{\mathcal{A}}_{ij}|$, also enforces non-negativity. While this avoids the noise-inflation pitfalls of $t_1$ and $t_2$, it inherently fails to distinguish between co-moving (positive) and hedging (negative) relationships. Although it preserves strong correlations (generating more trading signals than $t$ and potentially identifying a greater number of economic clusters), these abundant signals are fundamentally unreliable. Without differentiating the sign of the correlation, there is no guarantee that the diverging prices within a cluster will mean-revert. Thus, any advantage $t_3$ holds over $t$ relies entirely on unpredictable price dynamics rather than systematic StatArb.

\section{Dense subgraph search using GBS} \label{app:dss_with_gbs}



A deep connection between GBS and subgraph searches has been extensively established~\cite{arrazola2018using, bradler2018gaussian}. In this work, we frame this connection through the lens of FCGs $\mathcal{G}$, which are described by a symmetric adjacency matrix $\mathbf{\mathcal{A}}$. The matrix entries are defined as $\mathbf{\mathcal{A}}_{ij} = w_{ij}$ if stocks $i$ and $j$ are correlated with weight $w_{ij}$, and $0$ otherwise. Because this adjacency matrix is inherently symmetric, we can directly map the FCG to the input matrix of a GBS device and, by extension, to a Gaussian state. 

In this context, the hafnian functions as a weighted sum of perfect matchings. For non-negative graphs, the hafnian correlates positively with graph connectivity; highly connected graphs yield larger hafnians due to the accumulation of positive edge weights. This relationship is evident in Equation~(\ref{haf}), where the $B_{ij}$ terms provide positive contributions. Notably, this positive correlation persists even for general and complex-weighted graphs~\cite{deng2023solving}. Furthermore, the proportionality between graph connectivity and the number of perfect matchings has been thoroughly explored in~\cite{aaghabali2015upper}. Formally, the connectivity of a graph is quantified by its density:

\begin{align} \label{eqn:graph_density}
    D(\mathcal{G}) = \frac{2|E|}{|V|(|V| - 1)},
\end{align}

\noindent where $|E|$ and $|V|$ denote the number of edges and vertices in $\mathcal{G}$, respectively.

Consequently, GBS inherently samples from a probability distribution heavily biased toward dense subgraphs. For FCGs, we utilise weighted density, substituting $|E|$ in Equation~(\ref{eqn:graph_density}) with the sum of the edge weights, $\sum_{ij} \mathbf{\mathcal{A}}_{ij}$. Because the hafnian enumerates perfect matchings, one might expect the subgraphs sampled by GBS to always contain an even number of vertices. This aligns with lossless GBS dynamics, where photons generated from two-mode squeezed vacuum states strictly arrive in pairs. 

However, because we employ threshold detectors (registering only the presence or absence of photons), multiple photons arriving at a single mode are recorded as a single click. As a result, the recorded photon number statistic, and thus the number of vertices in the sampled subgraph, can be odd, despite a perfectly lossless system.

To harness this intersection of finance, graph theory, and GBS, we must encode the FCG $\mathcal{G}$ into the physical GBS device. First, we define the matrix $\mathbf{B}$ from Equation~(\ref{gbs_prob_main}) as $\mathbf{B} := c\mathbf{\mathcal{A}}$, where $c > 0$ is a scalar rescaling parameter. The standard graph encoding protocol for lossless GBS proceeds as follows~\cite{bromley2020applications}:

\begin{enumerate}
    \item Apply the Takagi-Autonne decomposition to the adjacency matrix $\mathbf{\mathcal{A}}$ to obtain
    $$ \mathbf{B} = \mathbf{U} \mathbf{\Lambda} \mathbf{U}^{\mathsf{T}}, $$
    where $\mathbf{U}$ is a unitary matrix and $\mathbf{\Lambda} = \bigoplus_{k=1}^m \lambda_k$ is a diagonal matrix of the singular values of $\mathbf{\mathcal{A}}$. These singular values $\lambda_k \in [0, 1)$ define the relationship between the squeezing parameters $r_k$, given by
    $$ r_k := \tanh^{-1}(\lambda_k), $$
    and the mean photon number $\bar{n}$ of the GBS distribution, defined as
    $$ \bar{n} := \sum_{k=1}^m \frac{\lambda_k^2}{1- \lambda_k^2}. $$
    \item Encode the unitary matrix $\mathbf{U}$ into the linear interferometer. This determines the appropriate beamsplitter and phase shifter configurations.
    \item Solve for the rescaling parameter $c > 0$ such that $\bar{n} = \sum_{k=1}^m \frac{(c\lambda_k)^2}{1-(c\lambda_k)^2}$. Alternatively, finding a $c$ that satisfies $0 < c < 1/\lambda_{\text{max}}$ (where $\lambda_{\text{max}}$ is the largest singular value of $\mathbf{\mathcal{A}}$) is also sufficient. This rescaling parameter is vital for encoding any arbitrary symmetric graph; it scales the adjacency matrix so that the new singular values $c\lambda_k$ strictly fall within $[0,1)$, thereby regulating both the squeezing and the overall mean photon number.
    \item Program the $k^{\text{th}}$ squeezed light sources using the updated squeezing parameters, ensuring $r_k = \tanh^{-1}(c\lambda_k)$.
\end{enumerate}

\noindent This procedure yields a GBS device that samples a photon number statistic $\mathbf{n}$ with the probability distribution:

\begin{align} \label{gbs_dist}
    \text{Prob}(\mathbf{n}) \propto c^N \frac{|\text{Haf}(\mathbf{B}_{\mathbf{n}})|^2}{\prod_{k=1}^m n_k!}.
\end{align}

\section{Photon loss models and squeezing approximations} \label{app:lossy_gbs}

Photon loss remains a primary obstacle to scaling photonic architectures and demonstrating near-term quantum advantage. To evaluate the robustness of our quantum clustering algorithms, we employ the standard beamsplitter loss model \cite{garcia2019simulating, oszmaniec2018classical}: 

\begin{align}
    \hat{a}_{\text{out}} = \sqrt{\eta}\,\hat{a}_{\text{in}} + \sqrt{1-\eta}\,\hat{b},
\end{align}

\noindent where $\hat{a}$ is the signal mode, $\hat{b}$ is the vacuum environment mode, and $\eta \in [0,1]$ is the transmissivity. This coupling attenuates the signal amplitude by $\sqrt{\eta}$ and injects vacuum noise, yielding a mixed state that degrades the underlying graph correlations. Because uniform loss is a Gaussian operation, the loss channel commutes with the linear interferometer. Therefore, it can equivalently be modelled at the input, manifesting as a reduction in the effective squeezing parameters.

For efficient classical simulation, this attenuation can be approximated by linearly rescaling the adjacency matrix: $\mathbf{\mathcal{A}} \to \eta\mathbf{\mathcal{A}}$ \cite{oh2024classical, zhao2025boosting}. This approach assumes sufficiently small squeezing (standard in most experimental implementations). Expressing the beamsplitter model in the momentum quadrature (setting $\hbar=1$) yields:

\begin{align} 
    \hat{p}_{\text{out}} = \sqrt{\eta}\,\hat{p}_{\text{in}} + \sqrt{1-\eta}\,\hat{p}_{\text{vac}},
\end{align}

\noindent Taking the variance of both sides recovers the exact identity linking the effective squeezing $s$ to the input squeezing $r$ \cite{oh2024classical}: $e^{-2s} = \eta e^{-2r} + 1 - \eta$. In the small squeezing limit ($r \ll 1$), this simplifies to $s \approx \eta r$. Consequently, the Takagi-Autonne decomposition becomes:

\begin{align} \label{eqn:ssq}
    \tilde{\mathbf{\mathcal{A}}} &= \mathbf{U} \bigoplus_k \tanh(s_k) \mathbf{U}^{\mathsf{T}} \approx \eta\mathbf{U} \bigoplus_k \tanh(r_k) \mathbf{U}^{\mathsf{T}} = \eta\mathbf{\mathcal{A}}.
\end{align}

\noindent Crucially, while GBS utilises multi-mode Gaussian states, this single-mode derivation remains robust; via the Bloch-Messiah reduction \cite{weedbrook2012gaussian, braunstein2005squeezing}, any multi-mode Gaussian state generated by linear optics decomposes into independent single-mode squeezers. 

The commutation of uniform loss channels with linear interferometers thereby permits the uniform rescaling of the adjacency matrix's singular values. Finally, we note a structural shift in the output: whereas lossless GBS (with photon number resolution) samples even-sized clusters due to the properties of the hafnian, photon loss breaks this constraint, allowing for clusters of arbitrary size parity.

We now formalise the assumption of sufficiently small squeezing utilised to approximate Equation (\ref{eqn:ssq}). Experimentally, small squeezing regimes are advantageous as they mitigate threshold detector collisions and higher-order instabilities, while easing state generation. We demonstrate that for the lossy configurations examined in Section \ref{sec:bench}, the approximation error remains bounded.

This error is defined by the difference between the effective adjacency matrix $\tilde{\mathcal{A}}$ and the scaled adjacency matrix $\eta\mathcal{A}$:

\begin{align} \label{app:eqn:mat_diff}
    \tilde{\mathcal{A}} - \eta\mathcal{A} = \mathbf{U}\underbrace{\bigoplus_k \bigl(\tanh(\eta r_k) - \eta \tanh(r_k)\bigr)}_{:=\mathbf{D}}\mathbf{U}^{\mathsf{T}}.
\end{align}

\noindent Taking the Frobenius norm and exploiting the unitary invariance of $\mathbf{U}$ yields:

\begin{align}
    \|\tilde{\mathcal{A}}-\eta\mathcal{A}\|_{F} = \|\mathbf{U}\mathbf{D}\mathbf{U}^{\mathsf{T}}\|_{F} = \|\mathbf{D}\|_{F}.
\end{align}

Assuming $r_k \ll 1$ for all $k$, we Taylor expand $\tanh(\cdot)$. For transmission efficiency $\eta \in [0,1]$:

\begin{align}
    \|\mathbf{D}\|_{F} &= \left\|\bigoplus_k \bigl(\tanh(\eta r_k) - \eta \tanh(r_k)\bigr)\right\|_F\\
    &= \left\| \bigoplus_k \left(\eta r_k - \frac{\eta^3 r_k^3}{3} - \eta r_k + \frac{\eta r_k^3}{3} +\mathcal{O}(r_k^5)\right)\right\|_F \\
    &\leq \sqrt{\sum_k \left(\frac{\eta(1-\eta^2)}{3}r_k^3\right)^2} \\
    &\leq \sum_k \left|\frac{\eta(1-\eta^2)}{3}r_k^3\right| \\
    &= \frac{\eta(1-\eta^2)}{3} \|r_k\|_3^3.
\end{align}

Thus, the approximation error is bounded by:

\begin{align} \label{app:eqn:err_approx}
    \|\tilde{\mathcal{A}} - \eta\mathcal{A}\|_F \leq \frac{\eta(1 - \eta^2)}{3} \|r_k\|_3^3 + \mathcal{O}(r_k^5).
\end{align}

\noindent For Section \ref{sec:loss}, Table \ref{tab:rescale_error} highlights two key observations. Firstly, significant error emerges for $\eta \in [0.2, 0.8]$. However, approximately 98\% of this error stems from the maximum squeezing parameter ($r_1 \approx 1.513$), which dominates the cubic and higher-order terms in Equation (\ref{app:eqn:err_approx}). Consequently, this dominant mode disproportionately inflates specific subgraphs sampled by GBS, analogous to how the market mode dominates a correlation matrix. 

Secondly, the error exhibits a non-monotonic trajectory with respect to $\eta$, peaking near $\eta \in [0.5, 0.6]$. At high loss (low $\eta$), both $\tilde{\mathcal{A}}$ and $\eta\mathcal{A}$ asymptotically approach the zero matrix, naturally diminishing their absolute difference.

\begin{table}[t!]
\centering
\setlength{\tabcolsep}{24pt} 
\renewcommand{\arraystretch}{1.3} 
\begin{tabular}{cc}
\hline
$\eta$ & Error \\
\hline
0.1 & 0.061 \\
0.2 & 0.116 \\
0.3 & 0.159 \\
0.4 & 0.185 \\
0.5 & 0.194 \\
0.6 & 0.184 \\
0.7 & 0.158 \\
0.8 & 0.118 \\
0.9 & 0.064 \\
1.0 & 0.000 \\
\hline
\end{tabular}
\caption{Error across 50 modes in the approximation of Equation (\ref{eqn:ssq}) for an $N=50$ stock universe across transmission rates $\eta$.}
\label{tab:rescale_error}
\end{table}

\section{Computational parameters and sample complexity concentration bound} \label{app:params}

Simulating quantum processes on classical hardware is computationally intensive, making sample acquisition from GBS a primary bottleneck. Within our StatArb strategy, this sampling process is invoked iteratively. For instance, over a sample period of $Y$ trading years (252 days/year) utilising an $l=3$ day window shift, we execute at least 84 clustering applications (provided the stop-win threshold $q$ is not prematurely triggered). By imposing a dynamic target cluster count $K$ per window, GBS must iterate $K-1$ times to extract batches of $H$ samples. When evaluating the strategy across $R$ backtest repetitions and $L$ loss rates, the total GBS calls scale to $Q=84YRLH(K-1)$. Given that classical hafnian calculations require $\mathcal{O}(N^3 2^N)$ time \cite{quesada2020exact}, this scaling $Q$ creates a severe simulation bottleneck even for conservative stock universes ($N = [20, 50]$).

Due to these simulation constraints, we restrict our sampling budget to $H = \mathcal{O}(N \log N)$. Although this is significantly smaller than a standard $\mathcal{O}(N^2)$ benchmark from a pure complexity standpoint, we demonstrate that this restricted budget captures enough of the high-probability heavy tail to generate competitive StatArb signals (yielding the advantageous lossless results in Section \ref{sec:bench}). When photon loss is introduced, the effective sample size drops because we must post-select the output, discarding any configurations with fewer than two distinct photon clicks. To maintain an adequate volume of valid samples for cluster determination, we scale the initial sampling budget by a factor $1/\eta$, where $\eta$ is the transmission efficiency.

In the ideal lossless regime, we calibrate the mean photon number to $\bar{n} = \sqrt{\dim(\mathcal{A})}$, which organically decreases as the quantum algorithms recursively extract subgraphs. Under physical loss, however, we must maintain a consistent $\bar{n}$ across varying loss rates to ensure balanced clustering. A naive compensation method is to uniformly increase the squeezing parameter; however, this predominantly amplifies thermal photons. 

As articulated in \cite{oh2024classical}, when sampling from a lossy covariance matrix, artificially inflating squeezing leaves quantum correlations static while uniformly amplifying classical noise. Consequently, the output distribution drifts from a quantum GBS distribution toward a thermal distribution, yielding subgraphs that are increasingly random and less dense. This degradation is corroborated by experimental implementations \cite{sempere2022experimentally}, demonstrating that lower squeezing values actually require fewer samples to successfully isolate dense subgraphs.

To preserve the target mean photon number without inducing thermal degradation, we instead compensate by injecting coherent light via displacement \cite{mer2026}. Let $\bar{n}_{\text{sq}}$ denote the initial mean photon number derived purely from squeezing. Under loss, this attenuates to $\bar{n}_{\text{sq}}^{(l)} = \eta\bar{n}_{\text{sq}}$, where $\eta = 1 - l_r \in [0,1]$. We restore this photon deficit by applying mode-specific displacement amplitudes $\alpha_i \in \mathbb{C}$, yielding a compensated total input:

\begin{align}
    \bar{n}_{\text{tot}}^{(c)} = \bar{n}_{\text{sq}}^{(l)} + \bar{n}_{\text{disp}}^{(l)} = \eta\left(\sum_i^m \sinh^2{r_i} + \sum_i^m |\alpha_i|^2\right).
\end{align}

\noindent We tightly constrain these displacement amplitudes to ensure the relative weights of the encoded residual correlations remain invariant before and after compensation. A brief overview of DGBS is provided in Appendix \ref{app:d_gbs}.

We next justify the sufficiency of restricting the sampling budget to $H=\mathcal{O}(N \log N)$ per GBS call by demonstrating that the probability of failing to sample a sufficiently dense subgraph is negligible for a specified \textit{hit rate}.

Let $\text{WD}_{\text{max}}$ denote the maximum possible weighted density of a subgraph within a FCG $\mathcal{G}$, and consider an arbitrary sample $S$. We define the $\varepsilon$-optimal hit rate as:

\begin{align}
    p_\varepsilon = P_{\text{GBS}}(\text{WD}(S) \geq \text{WD}_{\text{max}} - \varepsilon).
\end{align}

\noindent This gives the probability that an arbitrary subgraph achieves a weighted density within $\varepsilon$ of the global maximum. The clustering algorithm theoretically fails if all $H$ independent samples fall below this tolerance. Assuming i.i.d. GBS samples, the probability of $H$ consecutive failures is bounded by:

\begin{align} \label{app:eqn:ineq}
    (1 - p_\varepsilon)^H \leq e^{-H p_\varepsilon} \leq \delta.
\end{align}

\noindent Rearranging yields the sample bound $H \geq \frac{\log(1 / \delta)}{p_\varepsilon}$ for a target failure probability $\delta \in (0,1)$. Substituting $H=F N\log N$ into Equation (\ref{app:eqn:ineq}) yields:

\begin{align}
    \exp(-F N \log N \cdot p_\varepsilon) \leq \delta.
\end{align}

\noindent Although $p_\varepsilon$ cannot be analytically computed efficiently, literature establishes that GBS generates a heavy-tailed distribution for dense subgraphs \cite{arrazola2018using, Oh2024quantum-inspired}. Applying the ansatz $p_\varepsilon = \Omega(1/N)$ provides a maximum failure probability of:

\begin{align}
    \delta_{\text{max}} = \exp\left(-FN \log N \cdot p_\varepsilon\right) = N^{-F}.
\end{align}

\noindent Thus, $\delta_{\text{max}}$ decays inverse-polynomially with universe size, confirming within a Probably Approximately Correct (PAC) framework \cite{10.1145/1968.1972} that $H=\mathcal{O}(N \log N)$ samples are sufficient. Specifically, for the analysis in Appendix \ref{app:mos} (where $p_\varepsilon \sim 1/30$), only scalings $F \ge 1$ guarantee a theoretical failure probability below 5\%. This mathematically reduces confidence in the $F \in \{0.01, 0.1\}$ regimes, but the failure probabilities for the $H$s used in Section \ref{sec:bench} were between 3.98\% and 6.46\%

\section{Displaced GBS}\label{app:d_gbs}

Displaced Gaussian Boson Sampling (DGBS) extends standard GBS by introducing coherent displacement operations, enabling the encoding of graph loops via arbitrary diagonal matrix elements. The resulting Gaussian state at the interferometer output is:

\begin{align}
    |\psi\rangle = \mathbf{U} \left( \bigotimes_{k=1}^m \hat{D}(\alpha_k) \hat{S}(r_k) |0\rangle_k \right),
\end{align}

\noindent where $\hat{S}(r_k)$ is the squeezing operator, $\mathbf{U}$ is the interferometer, and $\hat{D}(\alpha_k) = \exp(\alpha_k \hat{a}^\dagger_k - \alpha_k^* \hat{a}_k)$ is the displacement operator for mode $k$ with amplitude $\alpha_k \in \mathbb{C}$.

Displacement fundamentally alters photon-number statistics. The probability of detecting a pattern $\mathbf{n} = (n_1, \dots, n_m)$ shifts from the standard hafnian to the loop hafnian of a generalised matrix $\tilde{\mathbf{B}}$:

\begin{equation}
    \text{Prob}(\mathbf{n}) = \frac{1}{\mathcal{Z}} \frac{|\text{lhaf}(\tilde{\mathbf{B}}_{\mathbf{n}})|^2}{\prod_k^m n_k!},
\end{equation}

\noindent where $\mathcal{Z}$ is a normalisation constant and $\tilde{\mathbf{B}}_{\mathbf{n}}$ is the submatrix formed by repeating the $i$-th row and column of $\tilde{\mathbf{B}}$ $n_i$ times. While the hafnian enumerates perfect matchings on loopless graphs, the loop hafnian enumerates matchings on looped graphs, with the diagonal elements of $\tilde{\mathbf{B}}$ dictated by $\alpha_k$.

By injecting coherent light, $\alpha_k$ can be tuned to counteract attenuation, stabilising the mean photon number to successfully extract clusters even in high-loss regimes. Conversely, in lossless scenarios, DGBS permits the encoding of cross-asset residual correlations (via squeezing) and individual asset features; for instance, stocks exhibiting high Sharpe ratios can be distinctly weighted by assigning them larger displacement amplitudes.

\begin{figure*}[t] 
\includegraphics[width=1\linewidth,height=1\textheight,keepaspectratio]{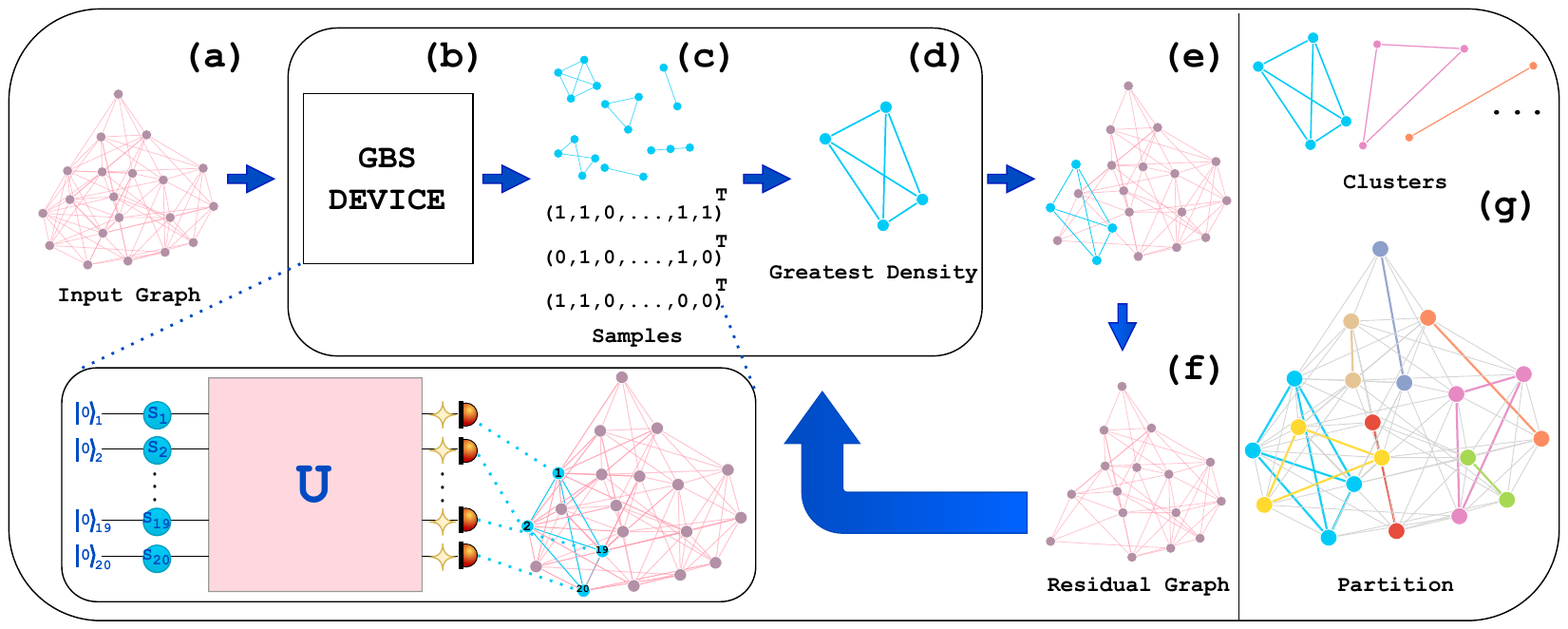}
\vspace*{-6.2cm}
\caption{The GBS Boost pipeline \cite{bonaldi2024boost}. (\textbf{a}) The initial FCG $\mathcal{G}$. (\textbf{b-d}) The iterative GBS sampling routine: (\textbf{b}) $\mathcal{G}$ is encoded into the GBS hardware (Section \ref{app:dss_with_gbs}), where single-mode squeezed vacuum states $\hat{S}(r_j)$ enter a linear interferometer defined by unitary $\mathbf{U}$. (\textbf{c}) Output threshold detectors yield photon-number statistics. Triggered modes correspond to specific nodes, sampling a subgraph of $\mathcal{G}$. (\textbf{d}) The algorithm evaluates these subgraphs, selecting the subgraph $\mathcal{S}$ with maximum density as the next cluster. (\textbf{e}) The nodes comprising $\mathcal{S}$ and their incident edges are excised to construct the residual graph $\mathcal{G}\setminus\mathcal{S}$. (\textbf{f}) This residual graph is fed back into the GBS routine to generate new samples. (\textbf{g}) This recursive extraction continues until the residual graph reaches size two (forming the final cluster), or size one at step $f+1$ (where the step $f$ residual becomes the final cluster), yielding a complete disjoint partition.}
\label{fig:boost_process}
\end{figure*}

\section{GBS Boost} \label{app:gbs_boost}

An important algorithm we adapt is the \emph{GBS-based clustering} framework (that we refer to as GBS Boost)  introduced by Bonaldi \textit{et al.} \cite{bonaldi2024boost}, modifying the initialisation to input our residual correlation matrix rather than a standard distance matrix. 

The algorithm operates as a greedy, iterative subgraph extraction protocol. Firstly, the GBS devices encodes a FCG $\mathcal{G}$ (with adjacency matrix $\mathcal{A}$), and produces samples. The sampled subgraph $\mathcal{S}_1$ exhibiting the highest density is extracted as the first cluster; ties are broken by selecting the subgraph with the maximum cluster value. The nodes of $\mathcal{S}_1$ are subsequently removed from $\mathcal{G}$. This sampling and extraction process recursively acts on the residual graph $\mathcal{G}_i = \mathcal{G}_{i-1} \setminus \mathcal{S}_i$ with updated adjacency matrix $\mathcal{A}_i$, terminating only when the residual graph can no longer be partitioned. Algorithm \ref{alg:boost} formalises this logic, and Figure \ref{fig:boost_process} visualises the pipeline.

\begin{algorithm}
\caption{GBS Boost \cite{bonaldi2024boost}}\label{alg:boost}
\begin{algorithmic}[1]
\Require Correlation matrix $\mathbf{\mathcal{C}}$ with graph $\mathcal{G}$, mean photon number $\bar{n}$, number of stocks $N$
\Ensure $C = \{C_1, C_2, \hdots\}$, a partition of $\mathcal{G}$
\State $C \gets \emptyset$
\State $\mathbf{\mathcal{A}} \gets \text{max}(0,\mathbf{\mathcal{C}} - \mathbf{I}) $ \Comment{GBS matrix} 
\While{$\mathbf{\mathcal{A}} \in \mathbb{R}_{[0,1]}^{N \times N}, N > 2$} \Comment{While $\mathcal{A}$ is large enough}
        \State $P \gets \texttt{GBS}(\mathbf{\mathcal{A}}, \bar{n}, N)$ \Comment{Get GBS samples $\{P_h\}_{h=1}^H$}
        \State $D \gets \{\text{D}(P_1), \hdots, \text{D}(P_H)\}$
        \State $\mathcal{S} \gets P_h \; \text{with} \; \arg\max(D)$
        
        \If{$N - |\mathcal{S}| = 1$} \Comment{Prevent singletons}
            \State $C \gets C \cup \mathcal{G}$ 
            \State \textbf{break}
        \Else
            \State $C \gets C \cup \mathcal{S}$
            \State $\mathbf{\mathcal{A}} \gets \text{reduce}(\mathbf{\mathcal{A}}, \mathcal{S})$ \Comment{$\mathcal{G} \gets \mathcal{G}\setminus \mathcal{S}$}
            \State $\bar{n}, N \gets \text{update}(\mathbf{\mathcal{A}})$ \Comment{Update parameters of $\mathbf{{\mathcal{A}}}$}
        \EndIf
    \EndWhile
\If{$N = 2$} \State $C \gets C \cup \mathcal{G}$ \Comment{Take residual graph as a cluster} 
\EndIf
\State \textbf{return} $C$
\end{algorithmic}
\end{algorithm}



On quantum hardware, GBS bypasses the primary computational bottleneck of classical graph clustering. Exact spectral methods, such as Spectral and SPONGE, are constrained by an $\mathcal{O}(N^3)$ eigendecomposition step, alongside $\mathcal{O}(MN)$ matrix construction and $\mathcal{O}(NK^2)$ $K$-means clustering \cite{zhugbs}. GBS circumvents this $\mathcal{O}(N^3)$ barrier by reframing clustering as probabilistic sampling. 

One-shot GBS feature extraction can achieve apparent linear scaling \cite{zhugbs}, GBS Boost and GBS Roots employ an iterative, greedy subgraph extraction protocol. Consequently, their runtime is governed by sequential GBS calls and the classical overhead of evaluating $H = \mathcal{O}(N \log N)$ samples per iteration and updating the residual adjacency matrix. Our hybrid pipeline successfully replaces deterministic $\mathcal{O}(N^3)$ diagonalisation with an efficient quantum sampling heuristic.

\section{Extended stock universe scaling analysis} \label{app:mos}

We first look at a small stock universe without loss and vary the magnitude of GBS samples, H. Larger stock universes generate larger FCGs. By restricting our sampling budget to $\mathcal{O}(N\log N)$ per GBS call, we risk under-sampling high-quality subgraphs in smaller universes. While necessary to circumvent computationally expensive runtimes, this restricted complexity ensures that the probability of sampling insufficiently dense subgraphs decays inverse-polynomially with the number of stocks. Consequently, this sample complexity achieves a lower theoretical error bound as the stock universe increases.

\begin{figure}[t]
\includesvg[width=0.9\linewidth,height=\textheight,keepaspectratio]{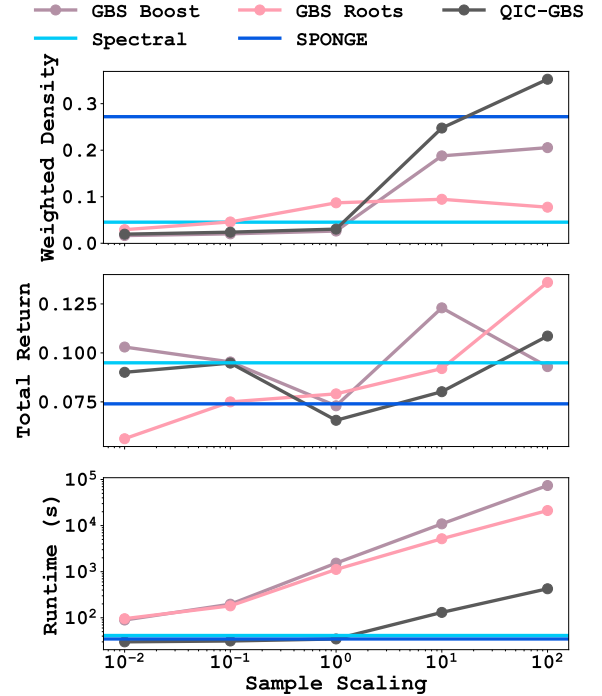}
\vspace*{-0.8cm}
\caption{Impact of the sample scaling factor ($F$) on average weighted density, total return, and runtime (seconds) for GBS Boost and GBS Roots ($N=30$). Scaling and runtime axes are base-10 logarithmic. Data spans $\sim$100 runs ($F \le 10^{-1}$), $\sim$50 runs ($F=10^0$), and $\sim$10 runs ($F \ge 10^1$). Horizontal benchmarks for Spectral and SPONGE denote 50 classical runs.}
\label{fig:mag_samp}
\end{figure}

Figure \ref{fig:mag_samp} evaluates a sample scaling factor $F \in \{10^{-2}, 10^{-1}, 10^0, 10^1, 10^2\}$, where the number of samples is $H = FN \log N$. For $N=30$, setting $F=10^1$ yields $H \approx N^2$. This represents the ideal sampling budget if simulation times were unbounded. We assess how this sample restriction impacts the quality of the sampled subgraphs and the subsequent returns generated.

In the upper panel, GBS Boost's weighted density sharply increases between $F = 10^0$ and $10^1$, whereas GBS Roots begins to plateau. The middle panel reveals steady return increases for GBS Roots, while GBS Boost plateaus near a 0.1 return. Crucially, both panels indicate that neither metric improves significantly beyond our ideal scale ($F=10^1$). Given the sub-exponential runtime growth conveyed in the bottom panel, exceeding $F=10^1$ provides no practical advantage for demonstrating competitiveness against Spectral and SPONGE on 30-stock datasets. Notably, $F=10^{-1}$ suffices to establish an economic edge over SPONGE.

At smaller scalings, QIC-GBS mirrors GBS Boost in both weighted density and total return, with runtimes comparable to the classical methods. At larger scalings, its runtime aligns with the lower-scale quantum methods while generating competitive returns. This establishes QIC-GBS as a highly efficient classical proxy while photonic quantum hardware matures.

Alternatively, a dynamic batch sampling procedure could mitigate low sample volumes without strictly inflating $F$. For instance, the procedure could collect up to $N$ batches of $H$ samples per GBS call, memorising the candidate with the highest weighted density. If this target metric does not improve over three consecutive batches, sampling terminates pre-emptively.

Now we scale the stock universe in single increments (20 to 30 stocks) and larger intervals (12 to 50 stocks) to analyse the impact on total return and Jaccard similarity. Because correlation structures shift as new assets are introduced, we do not expect monotonic performance trends. Furthermore, while stocks are added selectively to maintain GICS sector proportionality, smaller stratified subsets inevitably omit significant correlations

\begin{figure*}[ht!]
\includesvg[width=0.9\linewidth,height=0.85\textheight,keepaspectratio]{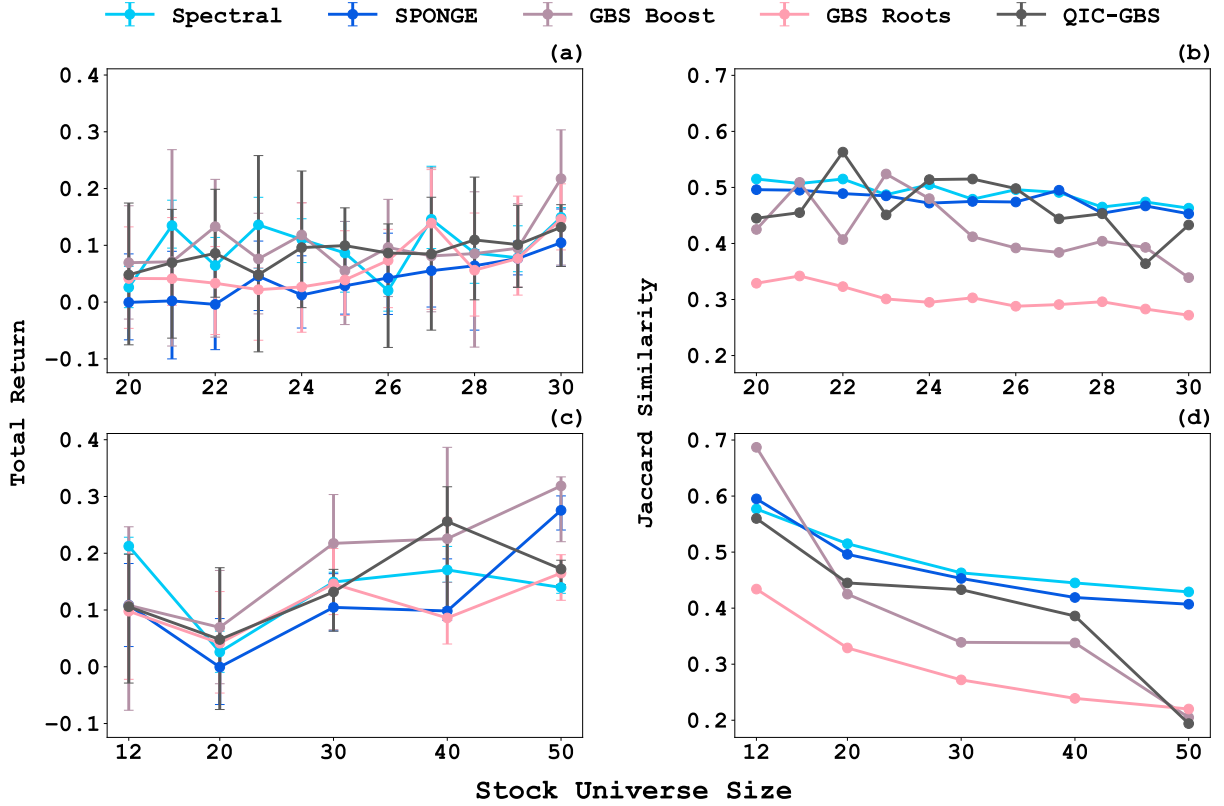}
\vspace*{-0.5cm}
\caption{Clustering performance across incrementally expanding stock universes. Left panels (a, c) display the median total return and IQR for each StatArb strategy as the universe scales from 20 to 30 stocks (a) and 12 to 50 stocks (c), retaining base constituent stocks across increments. Right panels (b, d) depict the corresponding average Jaccard similarities per stock universe. Quantum methods utilised $5N\log N$ samples per GBS call. Data spans 15--120 runs.}
\label{fig:scaling}
\end{figure*}

Figure \ref{fig:scaling}(a) and (c) indicate a general, albeit non-monotonic, upward trend in returns as the universe expands. Regarding total and risk-adjusted returns (Sharpe and Sortino ratios), the quantum methods remain highly competitive with their classical counterparts. In smaller universes ($N < 30$), classical algorithms demonstrate greater stability, evidenced by tighter interquartile ranges (IQRs), occasionally outperforming quantum methods that struggle with the idiosyncratic noise of small clusters. However, a performance transition occurs near $N=30$, where the lower return quartiles decouple from zero.


The primary advantage of the GBS methods manifests in return asymmetry. Classical algorithms exhibit narrow IQRs, indicating safety but limited upside. Conversely, at $N=50$, the upper quartile of GBS Boost nearly doubles that of Spectral, with its entire IQR sitting strictly above Spectral's. While GBS methods are intrinsically more volatile due to stochastic sampling and limited diversification in small clusters, their search for maximally dense subgraphs generates a heavy right tail in the return distribution. This increased variance is thus justified by significantly higher alpha potential in successful trials.

Figure \ref{fig:scaling}(b) and (d) illustrate that Jaccard similarity inversely scales with universe size, mirroring the large-scale lossless findings in Table \ref{table:lossless_clus}. Because larger graphs naturally accommodate greater clustering variation, classical methods consistently produce higher similarities than GBS methods. This corroborates the return profiles: deterministic classical methods identify dominant market modes, whereas GBS methods sacrifice inter-run similarity to discover high-return configurations.

Overall, incrementally expanding the stock universe yields non-monotonic return improvements across all algorithms. However, the decay in average Jaccard similarity suggests that inter-window cluster persistence degrades at larger scales, directly explaining the erratic return behaviour observed across these methods.

\section{Data and API} \label{app:data_api}

We source dividend-adjusted daily close prices from the Center for Research in Security Prices (CRSP) database \cite{wrds}. To isolate idiosyncratic stock behaviour, we extract the excess market return from the Fama-French 3-factor dataset \cite{wrds} to construct our single-factor residual returns. The complete dataset spans 465 S\&P 500 constituents from January 2004 to December 2024, but the main simulations are conducted over the volatile year 2020.

This temporal restriction, alongside a maximum universe cap of 100 stocks, is necessary to circumvent the prolonged GBS simulation runtimes. To ensure these constrained universes remain structurally representative, we construct stratified subsets based on the previous year's top-performing assets. Stocks are mapped to their appropriate sectors via Standard Industrial Classification (SIC) codes, allowing us to enforce proportional representation across the Global Industry Classification Standard (GICS) sectors: Energy, Materials, Industrials, Consumer Discretionary, Consumer Staples, Healthcare, Financials, Information Technology, Communication Services, Utilities, and Real Estate.

We simulate the quantum clustering methods using the Strawberry Fields API \cite{strawberryfields}. Specifically, we utilise the \texttt{sf.apps.sample.sample()} function, providing an input symmetric matrix $\mathcal{A}$, the number of samples $M$, and a constant mean photon number $\bar{n}=\sqrt{\dim(\mathcal{A})}$ across all loss rates. We employ threshold detection, recording a binary click if at least one photon occupies a given mode.

Consequently, the output detection probabilities are mathematically governed by the Torontonian matrix function rather than the hafnian \cite{quesada2018gaussian}. However, numerical relationships established via Monte Carlo simulations \cite{deng2023solving} confirm that the Torontonian closely tracks the hafnian for identifying dense subgraphs, validating our threshold-based heuristic. Formally, for a threshold click pattern $\mathbf{\tilde{n}} = \{\tilde{n}_1, \hdots, \tilde{n}_m \}$ and letting $\mathcal{N}$ denote the set of all underlying photon-number patterns yielding $\mathbf{\tilde{n}}$ (where at least one mode contains multiple photons), the Torontonian expands as:

\begin{align}
    \text{Tor}(\mathbf{I}-\mathbf{\tilde{\Sigma}}^{-1}) = \text{Haf}(\mathbf{B}_{\textbf{n}}) + \sum_{\tilde{n}\in\mathcal{N}} \frac{\text{Haf}(\mathbf{B}_{\mathbf{\tilde{n}}})}{\displaystyle \prod_{k=1}^m \tilde{n}_k!}.
\end{align}

\noindent Crucially, although exact classical simulation of threshold detection is \#P-hard and generally intractable for large systems, our parameter scaling ensures feasibility. The Walrus library evaluates matrices with an exponential time complexity that scales with the number of detected photons (submatrix dimension), rather than the total modes $N$. By bounding $\bar{n}=\sqrt{N}$, our largest simulated universe ($N=100$) is strictly constrained to $\bar{n}=10$. This restricts the Torontonian submatrices to computationally manageable sizes.

Finally, physical photon loss is modelled by passing the parameter \texttt{loss} $= l_r \in [0,1]$ into the sampling function, uniformly attenuating an $l_r$ fraction of the total photons within the GBS device.

\section{Metrics} \label{app:metrics}

We benchmark our algorithmic frameworks using both financial and graph-theoretic metrics. Financial performance is evaluated via total return (TR), Sharpe ratio (ShR), and Sortino ratio (SoR). The total return represents the difference between the final and initial value of a portfolio over the sample period. Because our temporal window spans exactly one year, TR serves as our baseline measure of raw economic profitability, functioning identically to the standard annualised return metric used in financial literature.  

To rigorously evaluate risk-adjusted performance, we utilise the Sharpe ratio, which penalises excess volatility. For a portfolio with returns $R_{t,p}$ over the evaluation period, expected portfolio return $\mathbb{E}[R_p] = \bar{R}_{t,p}$, and return standard deviation $\sigma_p$, it is defined as:

\begin{align}
    \text{ShR} = \frac{\bar{R}_{t,p} - R_f}{\sigma_p},
\end{align}

\noindent where $R_f$ represents the risk-free rate. A higher Sharpe ratio indicates stronger risk-adjusted performance. Similarly, the Sortino ratio replaces $\sigma_p$ with the downside deviation, penalising only the variability of negative returns. This provides a robust assessment of strategy safety during severe market drawdowns, making it a particularly appropriate metric for the volatile 2020 sample period.

Graph-theoretic metrics include weighted density, cluster value, and Jaccard similarity \cite{jaccard1901etude}. Weighted density (Equation \ref{eqn:graph_weighted_density}) assesses intra-cluster correlation strength relative to size, capturing the multi-stock co-movement that is fundamental to generating mean-reverting StatArb signals. Conversely, the cluster value (Equation \ref{value_func}) is a holistic metric that rewards total correlation cohesion across an entire partition. While weighted density evaluates individual cluster quality, the value metric evaluates the global clustering structure, naturally biasing towards larger clusters. Maximising this global value promotes the construction of highly cohesive, and profitable trading portfolios.

Finally, we utilise average Jaccard similarity to quantify cluster stability across consecutive re-balancing windows, which helps identify whether an algorithm produces stable or unstable clusters under shifting market regimes. At time $t$, the algorithm generates a partition

\begin{align}
    C = \{C_{t,i}\}_{i=1}^{K_t},    
\end{align}

\noindent of the FCG. When the rolling window shifts $l=3$ days forward, triggering a portfolio re-balance, a new partition $\tilde{C} = \{C_{t+l,i}\}_{i=1}^{K_{t+l}}$ is formed. We evaluate the overlap between these sets using the standard Jaccard index $J(A,B) = \frac{|A \cap B|}{|A \cup B|}$. For each new target cluster $C_{t+l,i} \in \tilde{C}$, we identify the maximally overlapping historical cluster $C_{t,j} \in C$:

\begin{align}
    \mathcal{J}(C_{t+l,i}) = \max_{j \in \{1, \dots, K_t\}} J(C_{t+l,i}, C_{t,j}).
\end{align}

\noindent The overall Jaccard similarity $S_{t+l}$ for the transition interval is defined as the mean of these maximal overlaps across all clusters in the current partition $\tilde{C}$:

\begin{align}
    S_{t+l} = \frac{1}{K_{t+l}} \sum_{i=1}^{K_{t+l}} \mathcal{J}(C_{t+l,i}).
\end{align}

\end{document}